\documentclass[12pt]{article}

\usepackage[margin=1in]{geometry} 
\usepackage{setspace}			

\usepackage{authblk}			
\usepackage{natbib}			
\usepackage{amsmath,amssymb}	
\usepackage{longtable}

\usepackage{booktabs}			
\usepackage{multirow}			
\usepackage{array}				
\usepackage{bm}
\usepackage{graphicx}			
\graphicspath{{./Figures}}		
\usepackage{float}				
\usepackage[section]{placeins} 	
\usepackage{caption}			
\usepackage{subcaption}			

\usepackage{caption}
\usepackage{url}				

\usepackage{xcolor}				
\usepackage[colorlinks=true,linkcolor=blue,citecolor=blue,urlcolor=blue]{hyperref}
\usepackage[normalem]{ulem}
\usepackage{hyperref}
\usepackage[nameinlink,noabbrev]{cleveref}

\usepackage{booktabs}
\usepackage{float}

\title{Portfolio Optimization and Tail-Risk Analytics of Actively Managed ETFs}

\author[1]{William W. Lamptey\thanks{Corresponding author, wilampte@ttu.edu}}
\author[1]{Nicholas Appiah}
\author[2]{Abootaleb Shirvani}
\author[1]{Priscilla Ati-Tay}
\author[1]{Svetlozar T. Rachev}
\author[3]{Frank J. Fabozzi}
\affil[1]{Department of Mathematics \& Statistics, Texas Tech University, Lubbock, TX,USA}
\affil[2]{Department of Mathematical Science, Kean University, Union, NJ, USA}
\affil[3]{Carey Business School, Johns Hopkins University, Baltimore, MD, USA}
\begin{document}
\date{}	
\maketitle

\begin{abstract}
This paper examines portfolio optimization and tail-risk analytics for a heterogeneous universe of actively managed investment funds. Using daily Bloomberg data for 30 funds from 4 December 2020 to 24 December 2025, the study evaluates buy-and-hold, mean--variance, CVaR-based, and tangency-type portfolio strategies under long-only and long--short constraints. The sample consists predominantly of actively managed exchange-traded funds, with PTTRX retained as an actively managed fixed-income mutual-fund comparator because of its relevance to the broader allocation setting. The empirical evidence shows substantial heterogeneity across thematic equity, fixed-income, income-oriented, multi-asset, and alternative strategies. This heterogeneity creates diversification opportunities, but also produces meaningful differences in volatility, drawdown behavior, downside exposure, and tail risk. Historical portfolio results show that tangency-type portfolios are generally the strongest competitors to the buy-and-hold benchmark in cumulative and risk-adjusted performance terms, whereas minimum-variance and CVaR-minimizing portfolios tend to sacrifice upside participation in exchange for stronger downside control. Dynamic allocation does not improve all strategies uniformly. In the long-only setting, the dynamic CVaR-95 portfolio is the most consistently attractive across several risk-adjusted criteria, while in the long--short setting, dynamic tangency-CVaR portfolios perform strongly but are more sensitive to turnover and implementation costs. Tail-risk diagnostics based on empirical VaR, Expected Shortfall, maximum drawdown, left-tail Hill estimators, and POT--GPD methods show that downside tail exposure remains meaningful after portfolio aggregation and varies across optimization rules, constraint regimes, and rebalancing specifications. Overall, the findings suggest that actively managed ETFs are best evaluated as components of a joint investment opportunity set in which dependence structure, portfolio design, dynamic allocation, implementation frictions, and tail-risk exposure jointly shape performance.
\end{abstract}


\section{Introduction}
\label{secIntroduction}

The debate between active and passive investment management remains one of the central questions in financial economics. At its core is whether professional portfolio managers can generate meaningful value after accounting for fees, risk exposure, and benchmark performance. Classic studies on mutual fund performance generally provide limited evidence of persistent net outperformance, documenting weak stock-selection ability or broad underperformance once costs are considered \citep{Sharpe1966,Jensen1968,BlakeEltonGruber1993,Malkiel1995,Gruber1996,Carhart1997,French2008}. At the same time, other studies suggest that active management may create value when information frictions, specialization, managerial skill, or portfolio concentration are important \citep{Ippolito1989,GrinblattTitman1989,GrinblattTitman1993,KacperczykSialmZheng2005,CremersPetajisto2009}. This unresolved tension continues to motivate empirical research on managed investment vehicles.

The growth of exchange-traded funds (ETFs) has reshaped this debate by changing how diversified investment strategies are packaged, traded, and implemented. ETFs combine exchange-based trading, intraday liquidity, operational flexibility, and a creation--redemption mechanism that helps align market prices with underlying net asset values \citep{Agapova2011,HuMorley2018,Haeberle2021}. Although ETFs were initially associated mainly with passive index-tracking products, the ETF wrapper has increasingly become a vehicle for active management. Actively managed ETFs therefore occupy an important position between traditional active mutual funds and passive ETFs by combining discretionary portfolio management with the trading and operational features of the ETF structure.

This structure creates both opportunities and challenges. Unlike passive ETFs, actively managed ETFs do not simply replicate a publicly observed benchmark. Their viability depends on balancing managerial discretion and proprietary information against the transparency needed for arbitrage, liquidity, and price efficiency \citep{HuMorley2018,Haeberle2021}. Recent regulatory and market developments have expanded the design space for active ETF products, making the segment increasingly relevant for investors, asset managers, and researchers \citep{HuMorley2018,Haeberle2021,AikenNesmithSchonberger2022}. As active ETFs expand across equity, fixed-income, multi-asset, and alternative investment mandates, the relevant question is no longer only whether individual active ETFs outperform passive benchmarks. It is also whether a diversified set of active ETFs can improve portfolio outcomes through diversification, downside-risk control, and exposure to distinct return drivers.

The empirical evidence on active ETFs remains mixed. Prior studies document variation in performance, tracking behavior, risk-adjusted returns, and the extent to which active ETFs deliver economically meaningful benefits relative to passive alternatives \citep{Rompotis2009,Rompotis2011,Rompotis2013,Dolvin2014,Schizas2014,Rompotis2015,BeckChongPhillips2017,Rompotis2022}. Subsequent work emphasizes heterogeneity across active ETF categories, fund size, investment style, and market segment. For example, \citet{GarynTal2013} identify an active ETF strategy with positive risk-adjusted excess returns, while \citet{BeckChongPhillips2017} show that large active ETFs may be more useful as components of broader asset-allocation problems than as isolated stand-alone holdings. Similarly, \citet{Meziani2015} argues that active ETFs may be most viable in relatively opaque or less efficient segments, particularly fixed income, where managerial discretion may be more valuable. More recent evidence suggests that active ETFs have attracted increasing investor interest, but often struggle to deliver clear abnormal performance under standard benchmark-based tests \citep{Rompotis2022,HilliardLe2021}.

A related literature asks whether actively managed ETFs are truly active. This issue matters because a fund may be classified as active while still closely tracking broad market or factor exposures. \citet{HilliardLe2021} show that actively managed ETFs do not always display substantially greater activeness than passive peers, and that performance varies across fund categories. Research on side-by-side management further shows that active mutual funds and active ETFs managed by the same sponsor may interact strategically through product design, tax treatment, investor clientele, and liquidity differences \citep{SherrillUpton2018,AikenNesmithSchonberger2022}. More broadly, ETFs are now used not only as end-investment vehicles, but also as implementation tools within active management itself \citep{Chen2024}. At the market level, ETF activity may affect liquidity, volatility, co-movement, price discovery, and informational efficiency in underlying securities \citep{BenDavidFranzoniMoussawi2018,LiebiEtAl2020,GlostenNallZou2021,KhomynRiddioughShim2024}. These issues matter for active ETF portfolio construction because they affect cross-fund dependence, diversification potential, drawdown behavior, and tail-risk exposure.

This paper contributes to the literature by shifting the analysis of actively managed ETFs from a narrow fund-level or benchmark-relative perspective to a portfolio-construction and tail-risk-management framework. If active ETFs are used as components of a broader investment opportunity set, their economic value cannot be assessed solely through stand-alone alpha or recent return performance. A fund that appears weak in isolation may still improve portfolio outcomes if it contributes diversification, reduces drawdown exposure, or changes the lower-tail behavior of the aggregate portfolio. Conversely, a fund with strong stand-alone performance may worsen portfolio tail risk if its returns are highly correlated with other risky exposures during adverse market conditions.

The portfolio framework used in this study is connected to the broader literature on risk-aware portfolio optimization. The classical mean--variance framework of \citet{Markowitz1952} remains the foundation of modern portfolio theory, but its sensitivity to estimation error and parameter uncertainty is well documented \citep{Michaud1989}. These limitations have motivated alternative optimization criteria that focus more directly on downside risk and extreme losses. In particular, the coherent risk-measure framework of \citet{ArtznerEtAl1999} and the conditional value-at-risk optimization approach of \citet{RockafellarUryasev2000,RockafellarUryasev2002} provide a natural foundation for portfolio construction when return distributions are asymmetric, heavy-tailed, or drawdown-sensitive. These concerns are especially relevant for actively managed ETFs because their exposures may vary through time and may differ substantially across equity, fixed-income, income-oriented, multi-asset, and alternative strategies.

We examine a universe of 30 actively managed investment funds over the period from 4 December 2020 to 24 December 2025. The sample consists predominantly of actively managed ETFs, with PTTRX retained as an actively managed fixed-income mutual-fund comparator because of its relevance to the broader allocation setting. The empirical analysis proceeds in four stages. First, we document fund-level behavior using cumulative price dynamics, return dependence, fund-size information, and summary statistics of daily returns. Second, we construct buy-and-hold, mean--variance, and conditional value-at-risk portfolios under both long-only and long--short constraints. Third, we compare historical full-sample optimization with a dynamic three-year rolling-window allocation design in which portfolio weights are updated through time using only prior information. Fourth, we evaluate downside and tail risk using empirical Value-at-Risk, Expected Shortfall, maximum drawdown, left-tail Hill estimators, and peak-over-threshold generalized Pareto diagnostics.

The paper makes three main contributions. First, it evaluates actively managed ETFs as interacting components of a joint portfolio opportunity set rather than only as stand-alone funds relative to passive benchmarks. Second, it compares mean--variance and CVaR-based optimization under both long-only and long--short constraints, allowing the roles of the risk measure and the admissible trading set to be examined directly. Third, it integrates conventional risk-adjusted performance analysis with downside-risk and extreme-tail diagnostics, showing that portfolios with similar Sharpe-ratio performance may differ materially in drawdown behavior, expected tail losses, and lower-tail thickness.

The empirical results show substantial heterogeneity across the actively managed fund universe. Some funds behave like aggressive growth or thematic exposures, others display lower-volatility income-oriented patterns, and selected alternative strategies provide diversification value through weaker or negative correlations. Portfolio optimization changes the investor experience in economically meaningful ways, but no single allocation rule dominates across all criteria. Tangency-type portfolios are generally the strongest competitors to the buy-and-hold benchmark in cumulative and risk-adjusted terms, whereas minimum-variance and CVaR-minimizing portfolios tend to sacrifice upside participation in exchange for stronger downside control. The dynamic results show that adaptive allocation does not improve all strategies uniformly. In the long-only setting, the dynamic CVaR-95 portfolio is the most consistently attractive across several risk-adjusted measures, while in the long--short setting, dynamic tangency-CVaR portfolios perform strongly across several reward-to-risk criteria. Tail-risk diagnostics further show that downside tail exposure remains meaningful after portfolio aggregation and varies across optimization rules, constraint regimes, and rebalancing specifications.

The remainder of the paper is organized as follows. Section~\ref{secData} describes the fund universe, sample construction, return definitions, summary statistics, cumulative price behavior, dependence structure, and fund-size characteristics. Section~\ref{secMethodology} presents the portfolio construction framework, including the buy-and-hold benchmark, mean--variance optimization, CVaR optimization, long-only and long--short constraints, historical and dynamic allocation designs, and performance evaluation methods. Section~\ref{secPortfolioResults} reports the historical and dynamic portfolio optimization results, risk-adjusted performance comparisons, bootstrap tests of Sharpe-ratio differences, and economic interpretation of the allocation mechanisms. Section~\ref{secTailRisk} examines downside-risk and tail-risk analytics, including empirical VaR, Expected Shortfall, maximum drawdown, Hill estimators, and POT--GPD diagnostics. Section~\ref{secConclusion} discusses practical implications, implementation issues, estimation-error concerns, and concludes the paper.

\section{Data and Preliminary Evidence}
\label{secData}

\subsection{Fund Universe and Sample Selection}
\label{subsecFundUniverse}

This study examines a universe of 30 actively managed investment funds over the period from 4 December 2020 to 24 December 2025. The sample consists predominantly of actively managed exchange-traded funds (ETFs), with PTTRX retained as an actively managed fixed-income mutual-fund comparator because of its relevance to the broader allocation setting. The resulting universe is therefore best interpreted as a predominantly active-ETF opportunity set supplemented by one intentionally retained fixed-income mutual-fund benchmark. The fund universe comprises AOA, AOK, AOM, AOR, ANGL, ARKF, ARKG, ARKK, ARKQ, ARKW, DBMF, DYNF, ELD, FBCG, GAA, GQRE, IVOL, JAAA, JCPB, JEPI, JMUB, JPST, KMLM, MUNI, PTTRX, PULS, QAI, RIGS, SRLN, and TOTL.

The funds were selected to provide broad coverage across the actively managed investment landscape while maintaining sufficiently long and continuous return histories for both historical and rolling dynamic portfolio analyses. The sample spans a wide range of investment styles, including thematic and innovation-oriented equity strategies, fixed-income and income-generating mandates, option-overlay and risk-managed approaches, alternative investment exposures, and diversified multi-asset allocations. This diversity is important because the objective of the study is not to evaluate a single category of active funds, but rather to examine how heterogeneous active investment vehicles interact within a common portfolio construction framework. By incorporating funds with distinct risk exposures, return drivers, and sensitivity to changing market conditions, the sample provides a representative environment for studying diversification, portfolio optimization, dynamic allocation, and downside-risk management.

Although the sample does not encompass the entire actively managed ETF universe, it captures a broad cross-section of economically significant active investment strategies and is therefore well suited for analyzing the portfolio opportunity set available to investors. Nevertheless, as with any empirical study, the possibility of sample-selection effects cannot be completely eliminated, and the results should be interpreted within the context of the selected investment universe.

\subsection{Return Construction and Empirical Grouping}
\label{subsecReturnConstruction}

For each fund, adjusted closing prices are used so that distributions and other price adjustments are reflected in the return series where applicable. Let \(P_{i,t}\) denote the adjusted closing price of fund (i) on trading day (t). Daily continuously compounded returns are computed as
\begin{equation}
r_{i,t}=\ln\left(\frac{P_{i,t}}{P_{i,t-1}}\right),
\label{eqLogReturn}
\end{equation}
where \(r_{i,t}\) is the log return on fund (i) between trading days (t-1) and (t).

To compare funds with different nominal price levels, each adjusted price series is also transformed into an indexed cumulative price process with a base value of 100 at the start of the sample:
\begin{equation}
I_{i,t}=100\times \frac{P_{i,t}}{P_{i,0}},
\label{eqIndexedPrice}
\end{equation}
where \(P_{i,0}\) is the adjusted closing price of fund (i) on 4 December 2020. The indexed series allow the analysis to compare cumulative growth, drawdown severity, and recovery patterns across funds on a common scale.

The data are aligned on a common daily trading calendar, with U.S. market holidays and non-trading days excluded from the sample. This alignment ensures that return comparisons, correlation estimates, and portfolio optimization exercises are based on a consistent set of observations across all funds. Because the subsequent analysis depends on cross-fund dependence, portfolio weights, and downside-tail behavior, maintaining a common trading-day structure is necessary for both comparability and numerical stability.

In addition to using the full 30-fund universe, the descriptive analysis organizes funds into six empirical price-dynamics groups based on similarities in their indexed cumulative price paths. This grouping is not intended to replace formal portfolio optimization. Rather, it provides an initial visual and economic classification of the sample by highlighting funds with similar growth trajectories, drawdown behavior, volatility patterns, and recovery profiles. The grouping is therefore used as a descriptive device that helps motivate the portfolio-construction analysis that follows.
\begin{equation}
\left\{
\begin{array}{l}
\text{Group 1: } \{\text{AOA, AOR, ARKQ, DYNF, FBCG}\} \\[4pt]
\text{Group 2: } \{\text{AOM, ARKW, DBMF, GAA, QAI}\} \\[4pt]
\text{Group 3: } \{\text{AOK, ARKF, GQRE, JEPI, KMLM}\} \\[4pt]
\text{Group 4: } \{\text{JAAA, JPST, MUNI, PULS, RIGS}\} \\[4pt]
\text{Group 5: } \{\text{ANGL, ELD, JCPB, JMUB, SRLN}\} \\[4pt]
\text{Group 6: } \{\text{ARKG, ARKK, IVOL, PTTRX, TOTL}\}
\end{array}
\right.
\label{eqEmpiricalGroups}
\end{equation}

Group 1 contains stronger growth-oriented funds with relatively high cumulative appreciation. Group 2 combines mixed growth and diversifying strategies with uneven recovery paths. Group 3 is an intermediate hybrid group with dispersed cumulative behavior. Group 4 is the most stable low-volatility fixed-income cluster. Group 5 contains riskier spread-sensitive fixed-income funds with wider drawdowns. Group 6 is a drawdown-heavy mixed group combining distressed growth funds with more resilient defensive names.

\subsection{Summary Statistics}
\label{subsecSummaryStatistics}

Table~\ref{tabSummaryStats} reports summary statistics for the daily log returns of the 30 funds included in the analysis. The table includes the mean, standard deviation, skewness, and kurtosis of each return series. Additional distributional information, including the minimum, maximum, and selected quantiles of each fund return distribution, is reported in Appendix~\ref{appFundStats}, Table~\ref{tabReturnQuantiles}. These statistics provide an initial view of the heterogeneity present within the sample and help motivate the subsequent portfolio optimization and tail-risk analysis.

The results reveal substantial variation across funds in terms of both return and risk characteristics. Equity-oriented and thematic growth funds generally exhibit higher volatility and wider return ranges, while fixed-income and income-oriented funds display lower volatility and more stable return profiles. Several funds exhibit noticeable skewness and excess kurtosis, indicating departures from normality and suggesting the presence of asymmetric and heavy-tailed return behavior. These findings support the use of downside-risk measures and tail-sensitive optimization methods in addition to traditional mean--variance analysis.

The evidence of non-normality is particularly relevant for the CVaR optimization and tail-risk diagnostics developed later in the paper. Since variance alone may not fully capture the magnitude and frequency of extreme losses, the distributional characteristics reported in Table~\ref{tabSummaryStats} provide an important empirical justification for the use of coherent risk measures and extreme-value techniques.

\begin{table}[H]
\centering
\caption{Summary Statistics of Daily Log Returns}
\label{tabSummaryStats}
\scriptsize
\begin{tabular}{lrrrr}
\toprule
Fund & Mean & Std. Dev. & Skewness & Excess Kurtosis \\
\midrule
ARKK & 0.0009 & 0.0254 & 0.0824 & 2.8737 \\
ARKG & -0.0000 & 0.0274 & 0.1873 & 0.5072 \\
ARKF & 0.0012 & 0.0218 & 0.0236 & 3.6315 \\
ARKQ & 0.0015 & 0.0199 & 0.1496 & 3.8613 \\
ARKW & 0.0015 & 0.0226 & -0.1307 & 2.9551 \\
FBCG & 0.0010 & 0.0153 & 0.2561 & 10.4955 \\
GQRE & 0.0001 & 0.0092 & -0.2273 & 3.8902 \\
JEPI & 0.0001 & 0.0070 & 0.3191 & 27.6095 \\
JPST & 0.0000 & 0.0009 & -3.4698 & 12.4865 \\
JCPB & 0.0000 & 0.0032 & -0.4733 & 0.9523 \\
DYNF & 0.0009 & 0.0104 & 0.1481 & 11.7489 \\
PTTRX & 0.0001 & 0.0033 & -0.3076 & 1.0450 \\
TOTL & 0.0000 & 0.0034 & -0.2409 & 9.8285 \\
ANGL & 0.0000 & 0.0037 & -0.3191 & 12.7145 \\
JMUB & -0.0000 & 0.0021 & -1.2924 & 7.4841 \\
AOR & 0.0004 & 0.0059 & 0.1542 & 10.1095 \\
AOA & 0.0005 & 0.0074 & 0.2186 & 13.0478 \\
AOM & 0.0003 & 0.0046 & 0.1839 & 7.9473 \\
AOK & 0.0002 & 0.0039 & -0.1882 & 3.9066 \\
GAA & 0.0003 & 0.0062 & -0.5972 & 2.7341 \\
KMLM & -0.0002 & 0.0064 & -0.6791 & 3.4154 \\
DBMF & 0.0002 & 0.0068 & -0.7815 & 5.2813 \\
QAI & 0.0002 & 0.0044 & -1.8792 & 21.9288 \\
IVOL & -0.0001 & 0.0062 & 0.0061 & 6.1292 \\
JAAA & 0.0000 & 0.0013 & -2.7645 & 13.7278 \\
RIGS & 0.0000 & 0.0059 & 0.0262 & 4.9386 \\
SRLN & -0.0000 & 0.0024 & -0.4702 & 17.7265 \\
ELD & 0.0001 & 0.0068 & 0.0087 & 2.5674 \\
MUNI & -0.0000 & 0.0023 & -0.7326 & 12.7293 \\
PULS & 0.0000 & 0.0010 & -3.6272 & 13.3490 \\
\bottomrule
\end{tabular}
\end{table}

\subsection{Cumulative Performance, Dependence, and Fund Size}
\label{subsecPreliminaryEvidence}

\begin{figure}[h!]
\centering
\includegraphics[width=\linewidth]{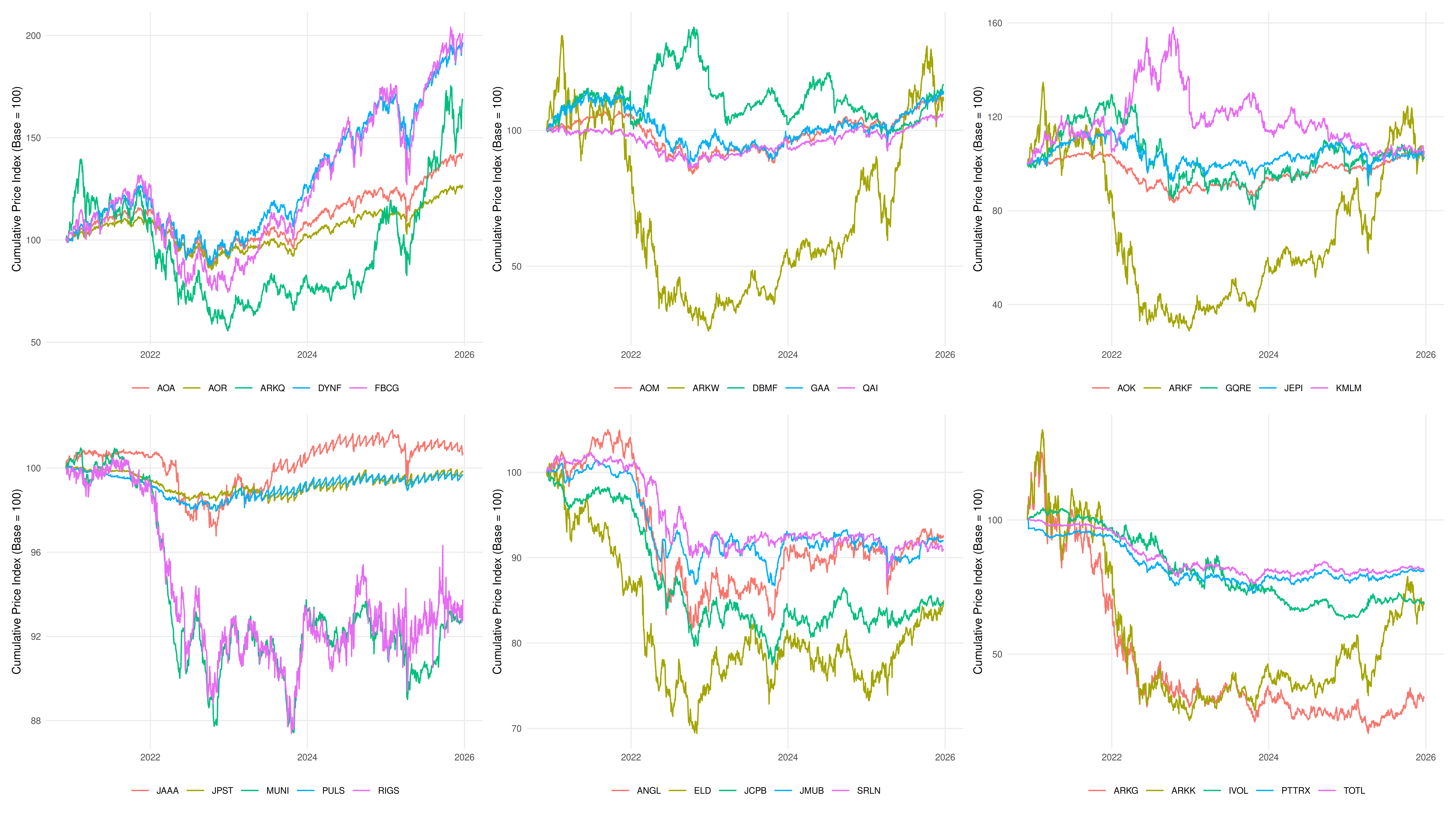}
\caption{Indexed cumulative price dynamics of the 30-fund universe grouped into six empirical price-dynamics clusters. Each series is normalized to a base value of 100 at the start of the sample on 4 December 2020. The panels highlight substantial cross-fund heterogeneity in cumulative growth, drawdown severity, and recovery profiles over the sample period ending 24 December 2025.}
\label{figCumPricePanels}
\end{figure}

Figure~\ref{figCumPricePanels} reports the indexed cumulative price dynamics of the 30 funds, grouped into the six empirical price-dynamics clusters defined in Equation~\eqref{eqEmpiricalGroups}. Each series is normalized to 100 at the beginning of the sample, allowing funds with different nominal price levels to be compared on a common scale. The figure shows substantial heterogeneity across the universe. Some funds display strong cumulative growth over the sample period, while others experience persistent drawdowns or much flatter cumulative paths. This dispersion is important because it suggests that the active ETF universe cannot be summarized by a single representative fund or by a simple average return pattern.

The upper panels of Figure~\ref{figCumPricePanels} contain several of the more growth-sensitive funds in the sample. DYNF and FBCG display especially strong cumulative appreciation, while ARKQ follows a more volatile path with a pronounced drawdown before a later recovery. DBMF also stands out because its trajectory is visibly different from many of the equity-sensitive funds, reflecting its role as a potential diversifying strategy. By contrast, ARKW and ARKF experience deep and persistent drawdowns before partial recovery. KMLM also displays a distinct path, with strong appreciation during parts of the sample and behavior that differs from the broad equity-oriented group.

The lower panels are more heavily populated by fixed-income and income-oriented funds. JAAA, JPST, and PULS exhibit comparatively stable cumulative paths, fluctuating within a much narrower range than the more volatile thematic growth funds. MUNI and RIGS show somewhat deeper drawdowns but remain less volatile than the most aggressive equity-oriented funds. ELD experiences a larger drawdown and slower recovery relative to other fixed-income-oriented funds, while ARKG and ARKK display some of the weakest cumulative paths in the sample. Overall, the cumulative price evidence indicates that the universe contains aggressive growth exposures, lower-volatility income-oriented products, and alternative or specialty strategies with more distinct return behavior.

\begin{figure}[h!]
\centering
\includegraphics[width=0.82\textwidth]{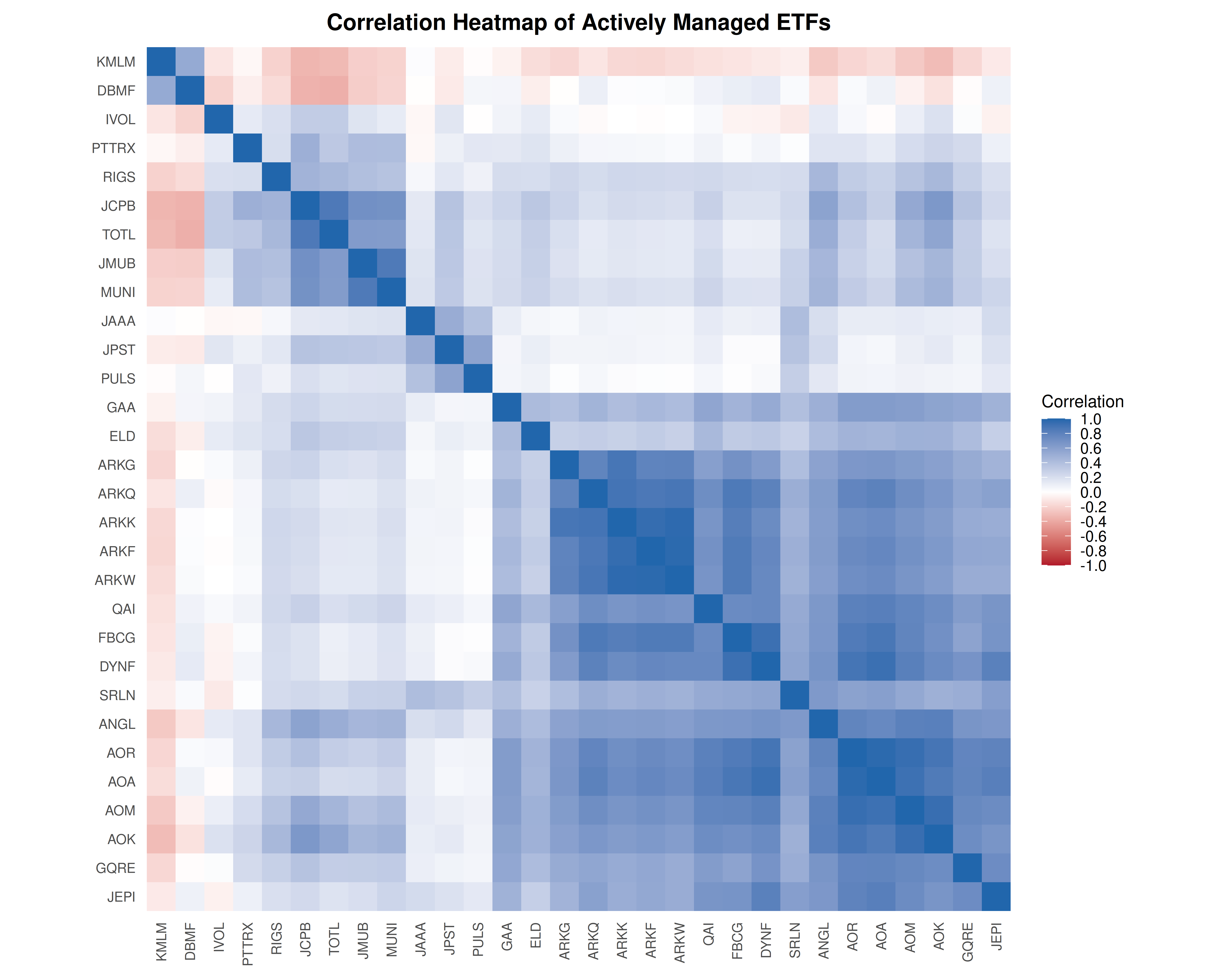}
\caption{Correlation heatmap of daily arithmetic returns for the 30 actively managed funds in the sample. Darker blue tones indicate stronger positive correlation, while red tones indicate negative correlation. The matrix reveals broad positive dependence across many equity, allocation, and hybrid funds, alongside weaker or negative correlation for selected specialty and alternative strategies.}
\label{figCorrHeatmap}
\end{figure}

Figure~\ref{figCorrHeatmap} reports the correlation matrix of daily arithmetic returns. The heatmap shows that dependence across the sample is predominantly positive, especially among equity, allocation, and hybrid funds. This broad positive dependence implies that many funds share common exposure to market-wide or growth-sensitive risk factors. At the same time, the correlation structure is not uniform. Several funds display weaker or negative correlations with large portions of the sample, suggesting the possibility of meaningful diversification benefits.

The strongest positive dependence appears among many of the equity-oriented, allocation, and hybrid strategies, including ARKG, ARKQ, ARKK, ARKF, ARKW, QAI, FBCG, DYNF, AOR, AOA, AOM, AOK, GQRE, and JEPI. These funds tend to move together more closely, indicating that diversification within this subset may be limited during common market movements. The heatmap also shows a more cohesive fixed-income block, especially among JCPB, TOTL, JMUB, and MUNI, with JPST and JAAA displaying weaker but still positive association with that group.

The most important diversification candidates are KMLM and DBMF. Both funds show low or negative correlation with many of the equity-sensitive funds, making them potentially valuable within optimized portfolios. IVOL also appears relatively weakly correlated with much of the universe, although its diversification role is less pronounced than that of KMLM and DBMF. These patterns suggest that diversification in this active-fund universe is unlikely to come simply from spreading weights evenly across all funds. Instead, it is more likely to come from combining growth-sensitive, income-oriented, and low-correlation alternative strategies in a deliberate allocation framework.

\begin{table}[h!]
\centering
\caption{Current Fund Size Snapshot: Net Assets / AUM}
\label{tabFundSize}
\footnotesize
\setlength{\tabcolsep}{3pt}
\renewcommand{\arraystretch}{1.08}
\begin{tabular}{@{}p{0.10\textwidth}p{0.60\textwidth}p{0.22\textwidth}@{}}
\toprule
Ticker & Fund & Net Assets / AUM (USD) \\
\midrule
AOA   & iShares Core 80/20 Aggressive Allocation ETF & 2.787B \\
AOK   & iShares Core 30/70 Conservative Allocation ETF & 0.742B \\
AOM   & iShares Core 40/60 Moderate Allocation ETF & 1.689B \\
AOR   & iShares Core 60/40 Balanced Allocation ETF & 3.240B \\
ANGL  & VanEck Fallen Angel High Yield Bond ETF & 2.980B \\
ARKF  & ARK Fintech Innovation ETF & 0.804B \\
ARKG  & ARK Genomic Revolution ETF & 1.296B \\
ARKK  & ARK Innovation ETF & 6.516B \\
ARKQ  & ARK Autonomous Technology \& Robotics ETF & 2.014B \\
ARKW  & ARK Next Generation Internet ETF & 1.596B \\
DBMF  & iMGP DBi Managed Futures Strategy ETF & 3.150B \\
DYNF  & iShares U.S. Equity Factor Rotation Active ETF & 30.112B \\
ELD   & WisdomTree Emerging Markets Local Debt Fund & 0.119B \\
FBCG  & Fidelity Blue Chip Growth ETF & 5.380B \\
GAA   & Cambria Global Asset Allocation ETF & 0.068B \\
GQRE  & FlexShares Global Quality Real Estate Index Fund & 0.342B \\
IVOL  & Quadratic Interest Rate Volatility and Inflation Hedge ETF & 0.491B \\
JAAA  & Janus Henderson AAA CLO ETF & 26.880B \\
JCPB  & JPMorgan Core Plus Bond ETF & 10.490B \\
JEPI  & JPMorgan Equity Premium Income ETF & 44.960B \\
JMUB  & JPMorgan Municipal ETF & 7.120B \\
JPST  & JPMorgan Ultra-Short Income ETF & 37.440B \\
KMLM  & KFA Mount Lucas Managed Futures Index Strategy ETF & 0.256B \\
MUNI  & PIMCO Intermediate Municipal Bond Active ETF & 2.764B \\
PTTRX & PIMCO Total Return Fund Institutional Class & 47.940B \\
PULS  & PGIM Ultra Short Bond ETF & 14.381B \\
QAI   & NYLI Hedge Multi-Strategy Tracker ETF & 0.787B \\
RIGS  & RiverFront Strategic Income Fund & 0.066B \\
SRLN  & State Street Blackstone Senior Loan ETF & 4.677B \\
TOTL  & State Street DoubleLine Total Return Tactical ETF & 4.110B \\
\bottomrule
\end{tabular}

\vspace{0.4em}
\parbox{0.94\textwidth}{\footnotesize Notes: Reported values are fund net assets / assets under management rather than equity-style market capitalization. The figures represent a current size snapshot rather than an end-of-sample December 2025 measure.}
\end{table}

Table~\ref{tabFundSize} reports a current snapshot of fund size using net assets or assets under management. The table shows wide variation in market footprint across the sample. PTTRX, JEPI, JPST, DYNF, PULS, JAAA, and JCPB are among the largest vehicles, indicating that the sample includes funds with substantial scale and investor adoption. At the other end, GAA, RIGS, ELD, KMLM, and GQRE are much smaller. This size dispersion matters because fund scale can affect liquidity, trading costs, market adoption, and the practical interpretation of portfolio allocation results.

Taken together, the preliminary evidence establishes three important facts. First, the actively managed fund universe is highly heterogeneous in cumulative price behavior, with large differences in growth, drawdown, and recovery patterns. Second, the dependence structure is mostly positive but not uniform, leaving room for diversification through selected low-correlation strategies such as KMLM and DBMF. Third, the funds differ substantially in size, suggesting that the sample includes both large, established vehicles and smaller niche strategies. These features motivate the portfolio optimization analysis that follows, since the value of the fund universe depends not only on individual performance but also on how the funds interact inside optimized portfolios.

\section{Portfolio Construction and Empirical Design}
\label{secMethodology}

This section describes the portfolio construction framework used to evaluate the actively managed fund universe. The analysis compares a passive buy-and-hold benchmark with optimized portfolios formed under mean--variance and conditional value-at-risk (CVaR) criteria. Both long-only and long--short constraint sets are considered, and the optimization exercise is implemented in both historical and dynamic settings. This design allows the paper to examine whether portfolio outcomes depend on the definition of risk, the admissible trading constraints, and the use of static versus adaptive allocation rules.

Let (N=30) denote the number of funds in the sample, and let
\[
\bm{r}_t=(r_{1t},r_{2t},\ldots,r_{Nt})^\top
\]
represent the vector of daily fund returns at time (t). Let   \(\bm{\mu}\in\mathbb{R}^{N}\) denote the vector of expected returns, and let \(\bm{\Sigma}\in\mathbb{R}^{N\times N}\) denote the covariance matrix of fund returns. For a portfolio with weight vector
\[
\bm{w}=(w_1,w_2,\ldots,w_N)^\top,
\]
portfolio variance is
\[
\sigma_p^2=\bm{w}^{\top}\bm{\Sigma}\bm{w},
\]
and portfolio volatility is
\[
\sigma_p=\sqrt{\bm{w}^{\top}\bm{\Sigma}\bm{w}}.
\]
The full-investment constraint requires
\begin{equation}
\bm{1}^\top\bm{w}=1.
\label{eqFullInvestment}
\end{equation}

\subsection{Buy-and-Hold Benchmark}
\label{subsecBuyHold}

The passive benchmark in the study is a buy-and-hold portfolio (BHP). This benchmark is initialized with equal weights across the 30 actively managed investment funds and is not rebalanced after formation. The initial allocation is therefore
\begin{equation}
w_{i,0}=\frac{1}{30}, \qquad i=1,\ldots,30.
\label{eqInitialBHPWeights}
\end{equation}
If \(R_{i,t}\) denotes the simple return of fund (i) during period (t), then the buy-and-hold portfolio return is
\begin{equation}
R^{\mathrm{BHP}}_{p,t}=\sum_{i=1}^{30}w_{i,t-1}R_{i,t}.
\label{eqBHPReturn}
\end{equation}
Because the portfolio is not rebalanced, the weights evolve endogenously according to relative fund performance:
\begin{equation}
w_{i,t}=
\frac{w_{i,t-1}(1+R_{i,t})}
{\sum_{j=1}^{30}w_{j,t-1}(1+R_{j,t})}.
\label{eqBHPWeightEvolution}
\end{equation}
This benchmark provides a simple passive reference point against which the optimized portfolios can be evaluated.

\subsection{Mean--Variance Optimization}
\label{subsecMeanVariance}

The first optimization framework is the classical mean--variance paradigm of \citet{Markowitz1952}, in which portfolio choice is based on expected return and variance, despite the well-known sensitivity of mean--variance optimization to estimation error \citep{Michaud1989}. For a given target return \(\mu_p\), the efficient portfolio solves
\begin{equation}
\min_{\bm{w}} \quad \bm{w}^{\top}\bm{\Sigma}\bm{w}
\label{eqMVEfficientObjective}
\end{equation}
subject to
\begin{equation}
\bm{\mu}^{\top}\bm{w}=\mu_p,
\qquad
\bm{1}^{\top}\bm{w}=1,
\label{eqMVEfficientConstraints}
\end{equation}
together with the relevant weight constraints. The resulting set of portfolios traces the mean--variance efficient frontier.

The global minimum-variance portfolio (MVP) is obtained by solving
\begin{equation}
\min_{\bm{w}} \quad \bm{w}^{\top}\bm{\Sigma}\bm{w}
\label{eqMVPObjective}
\end{equation}
subject to
\begin{equation}
\bm{1}^{\top}\bm{w}=1,
\label{eqMVPConstraint}
\end{equation}
and the relevant admissibility constraints. The MVP is therefore the portfolio with the lowest attainable variance within the feasible set.

When a risk-free rate \(r_f\) is available, the tangency portfolio (TVP) is defined as the portfolio that maximizes the Sharpe ratio:
\begin{equation}
\max_{\bm{w}} \frac{\bm{\mu}^{\top}\bm{w}-r_f}{\sqrt{\bm{w}^{\top}\bm{\Sigma}\bm{w}}}
\label{eqTangencyObjective}
\end{equation}
subject to
\begin{equation}
\bm{1}^{\top}\bm{w}=1.
\label{eqTangencyConstraint}
\end{equation}
This portfolio identifies the risky allocation with the highest excess return per unit of total volatility \citep{Tobin1958,Sharpe1966}.

\subsection{Conditional Value-at-Risk Optimization}
\label{subsecCVaROptimization}

The second optimization framework uses CVaR as the risk measure. CVaR is especially relevant in this setting because actively managed ETFs and related investment funds may display asymmetric, heavy-tailed, or drawdown-sensitive return behavior. Unlike variance, which treats upside and downside deviations symmetrically, CVaR focuses directly on expected losses beyond a specified tail threshold.

For confidence level \(\alpha\), the empirical CVaR objective can be written as
\begin{equation}
\zeta+
\frac{1}{(1-\alpha)T}
\sum_{t=1}^{T}
\max \left\{-\bm{w}^{\top}\bm{r}_t-\zeta,0\right\},
\label{eqCVaRObjectiveFunction}
\end{equation}
where (T) is the number of observations and \(\zeta\) is an auxiliary variable associated with the VaR threshold. The CVaR-minimizing portfolio solves
\begin{equation}
\min_{\bm{w},\zeta} \quad
\zeta+
\frac{1}{(1-\alpha)T}
\sum_{t=1}^{T}
\max\left\{-\bm{w}^{\top}\bm{r}_t-\zeta,0\right\}
\label{eqCVaRMinimization}
\end{equation}
subject to
\begin{equation}
\bm{1}^{\top}\bm{w}=1,
\label{eqCVaRFullInvestment}
\end{equation}
and the relevant weight constraints. In the empirical analysis, CVaR portfolios are constructed at the 95\% and 99\% confidence levels. These portfolios are denoted C95 and C99, respectively. The corresponding CVaR-based tangency-type portfolios are denoted TC95 and TC99.

Because all portfolio inputs are estimated from finite samples, the optimized weights may be sensitive to sampling variation. The empirical results should therefore be interpreted with the standard estimation-error caveat associated with mean--variance and downside-risk optimization.

\subsection{Long-Only and Long--Short Constraint Sets}
\label{subsecConstraints}

The optimization problems are solved under two constraint regimes. The first is a long-only constraint set, which requires all portfolio weights to be nonnegative:
\begin{equation}
w_i\geq 0, \qquad i=1,\ldots,30.
\label{eqLongOnlyConstraint}
\end{equation}
Together with the full-investment condition in Equation~\eqref{eqFullInvestment}, this constraint corresponds to a conventional fully invested allocation problem without short selling.

The second is a long--short constraint set. In this case, short positions are permitted, but individual weights are bounded to prevent extreme leverage or excessive concentration:
\begin{equation}
-\frac{1}{3}\leq w_i\leq 1.3, \qquad i=1,\ldots,30.
\label{eqLongShortConstraint}
\end{equation}
The full-investment condition in Equation~\eqref{eqFullInvestment} remains imposed. This constraint set expands the feasible allocation space relative to the long-only case and allows the optimization procedure to exploit relative-value positions across funds while still limiting extreme exposures \citep{JacobsLevyMarkowitz2005}.

\section{Portfolio Optimization Results}
\label{secPortfolioResults}

This section reports the historical and dynamic portfolio optimization results for the actively managed fund universe. The historical analysis provides a static full-sample benchmark for evaluating the risk--return trade-off of the 30-fund opportunity set, while the dynamic analysis examines whether rolling-window rebalancing improves performance when portfolio weights are updated through time using only prior information. The results are evaluated using cumulative wealth paths, efficient frontiers, risk-adjusted performance measures, bootstrap Sharpe-ratio tests, and transaction-cost robustness checks.

\subsection{Historical Portfolio Optimization Results}
\label{subsecHistoricalPortfolioResults}

\begin{figure}[htbp]
\centering
\includegraphics[width=\linewidth]{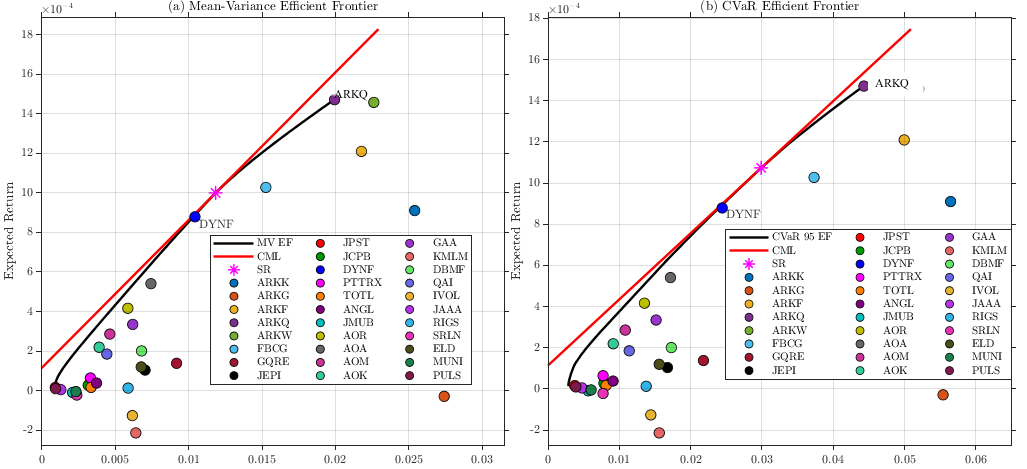}
\caption{Historical efficient-frontier results under mean--variance and CVaR portfolio optimization.}
\label{figMeanVarCVaR}
\end{figure}

Having established the descriptive properties of the actively managed fund universe in Section~\ref{secData}, we first examine whether portfolio optimization improves the risk--return trade-off relative to the buy-and-hold benchmark. Figure~\ref{figMeanVarCVaR} reports the historical efficient-frontier results under the mean--variance and CVaR optimization frameworks. The individual funds are widely dispersed in expected-return and risk space, which is consistent with the heterogeneity documented earlier. Several funds lie below the efficient frontier, indicating that diversified combinations of funds can dominate individual funds in isolation.

Panel~(a) of Figure~\ref{figMeanVarCVaR} presents the mean--variance frontier. The efficient frontier dominates most individual funds, while the capital market line identifies the reward-to-volatility trade-off associated with the tangency portfolio. Panel~(b) presents the corresponding CVaR frontier, where risk is measured by expected lower-tail losses rather than total return dispersion. The comparison shows that portfolio rankings depend on the definition of risk. Mean--variance optimization emphasizes the trade-off between expected return and volatility, whereas CVaR optimization places greater weight on downside losses. This distinction is especially relevant because several funds in the sample exhibit drawdowns, asymmetry, and tail-sensitive return behavior.

\begin{figure}[h!]
\centering
\includegraphics[width=\linewidth]{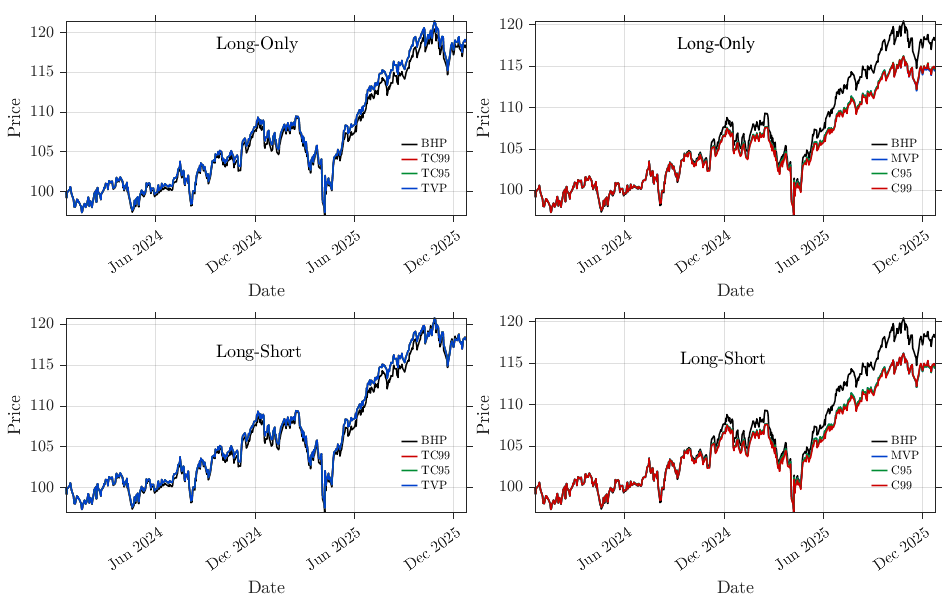}
\caption{Indexed cumulative wealth paths for the buy-and-hold benchmark and the optimized portfolios under long-only and long--short implementations.}
\label{figHistoricalCumulative}
\end{figure}

Figure~\ref{figHistoricalCumulative} reports indexed cumulative wealth paths for the buy-and-hold benchmark and the optimized portfolios under long-only and long--short implementations. All series are normalized to 100 at the beginning of the evaluation period. The long-only optimized portfolios broadly follow the direction of the buy-and-hold benchmark, consistent with the positive dependence structure observed in Figure~\ref{figCorrHeatmap}. Tangency-type portfolios are generally the strongest competitors to the benchmark in cumulative wealth terms, while minimum-variance and CVaR-minimizing portfolios display flatter paths because they prioritize volatility or downside-loss control rather than cumulative growth.

The long--short results are qualitatively similar. Although shorting expands the feasible set, the historical cumulative paths do not show a dramatic improvement relative to the long-only case. This suggests that, under the imposed long--short bounds, additional trading flexibility does not automatically generate superior realized wealth accumulation. Overall, the historical results show a clear trade-off: tangency-type portfolios are more competitive in growth terms, whereas minimum-variance and CVaR-minimizing portfolios provide more conservative allocations that sacrifice upside participation for stronger risk control.

\begin{figure}[h!]
\centering
\includegraphics[width=\linewidth]{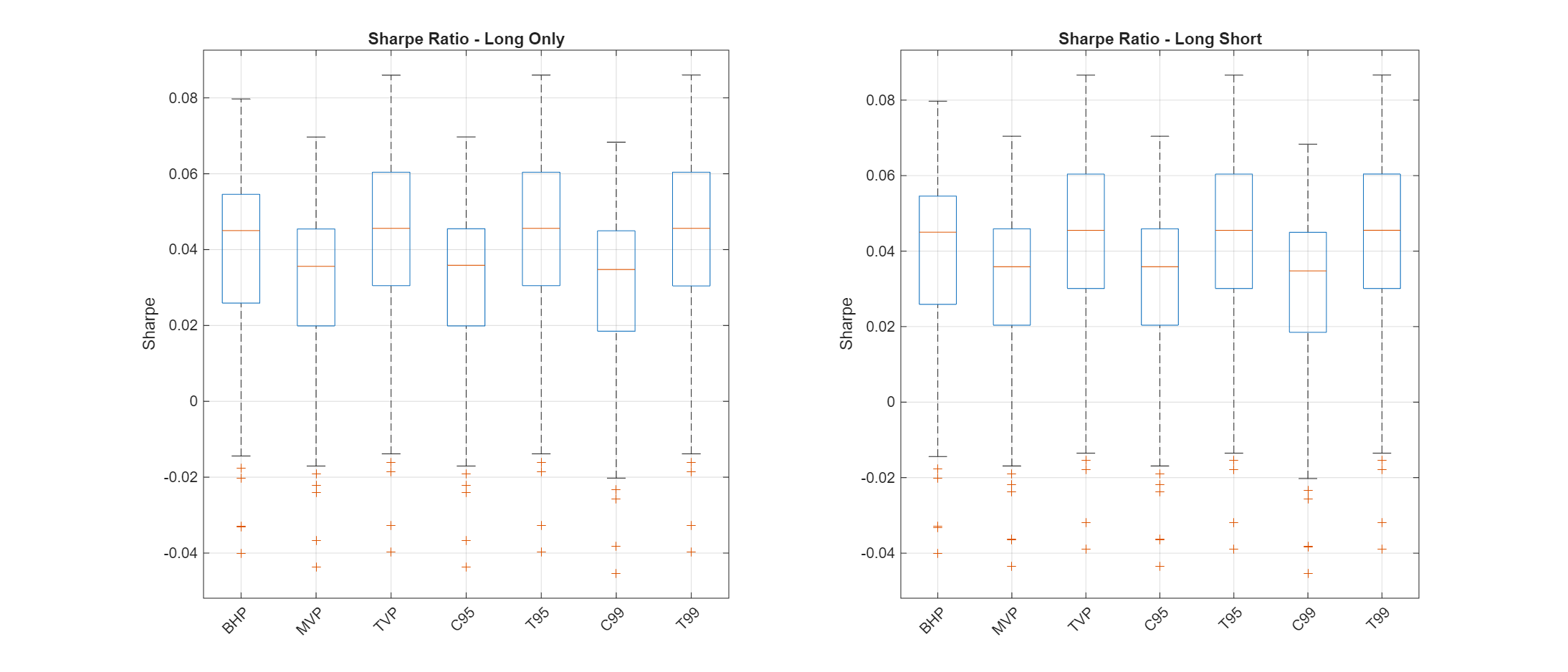}
\caption{Sharpe ratio distributions for the long-only and long--short portfolio strategies.}
\label{figSharpeLOLS}
\end{figure}

\begin{figure}[h!]
\centering
\includegraphics[width=\linewidth]{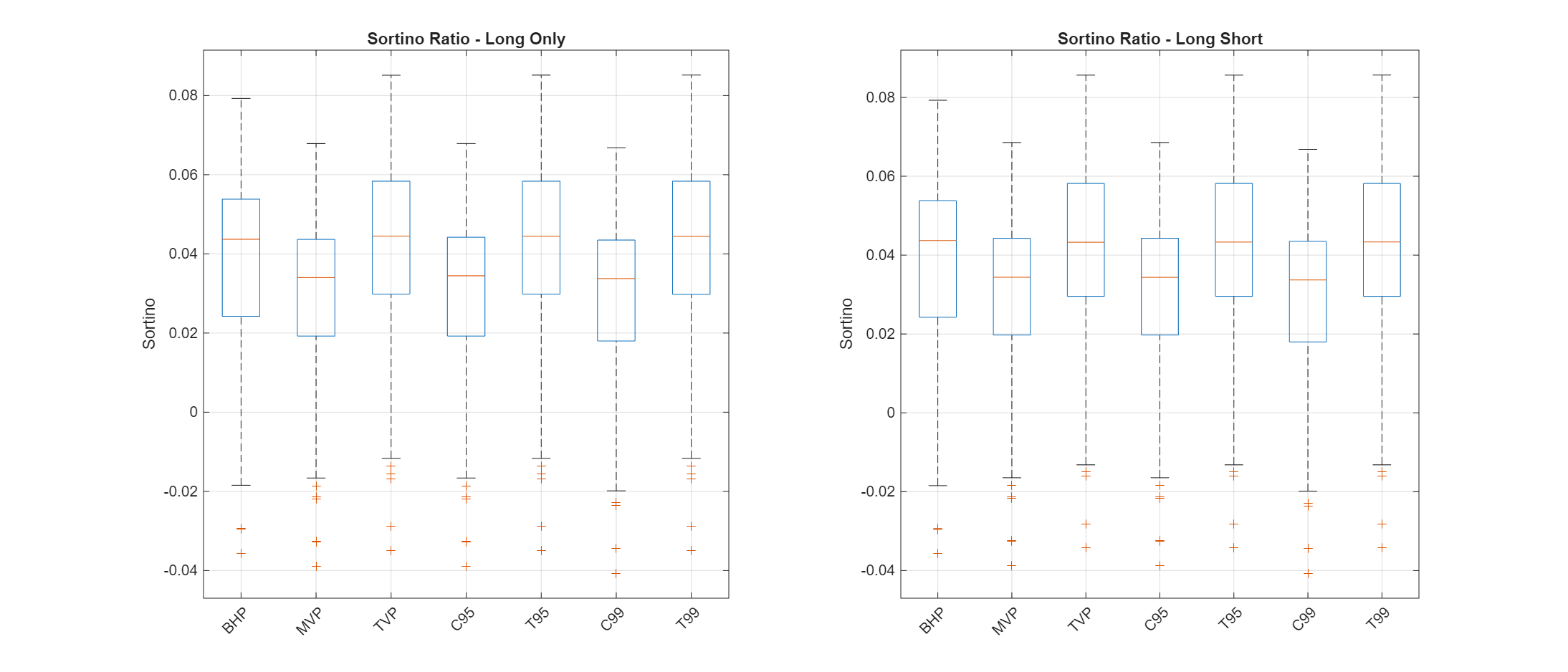}
\caption{Sortino ratio distributions for the long-only and long--short portfolio strategies.}
\label{figSortinoLOLS}
\end{figure}

\begin{figure}[h!]
\centering
\includegraphics[width=\linewidth]{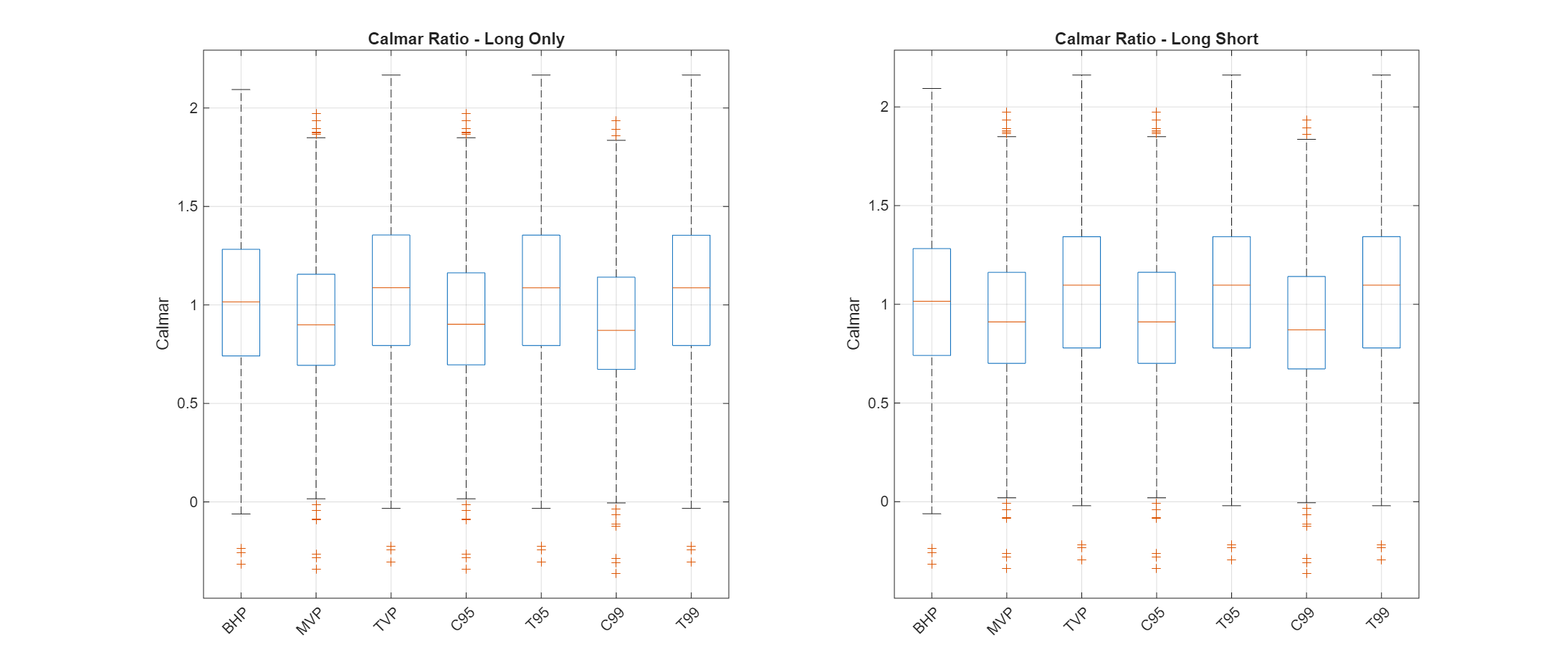}
\caption{Calmar ratio distributions for the long-only and long--short portfolio strategies.}
\label{figCalmarLOLS}
\end{figure}

\begin{figure}[h!]
\centering
\includegraphics[width=\linewidth]{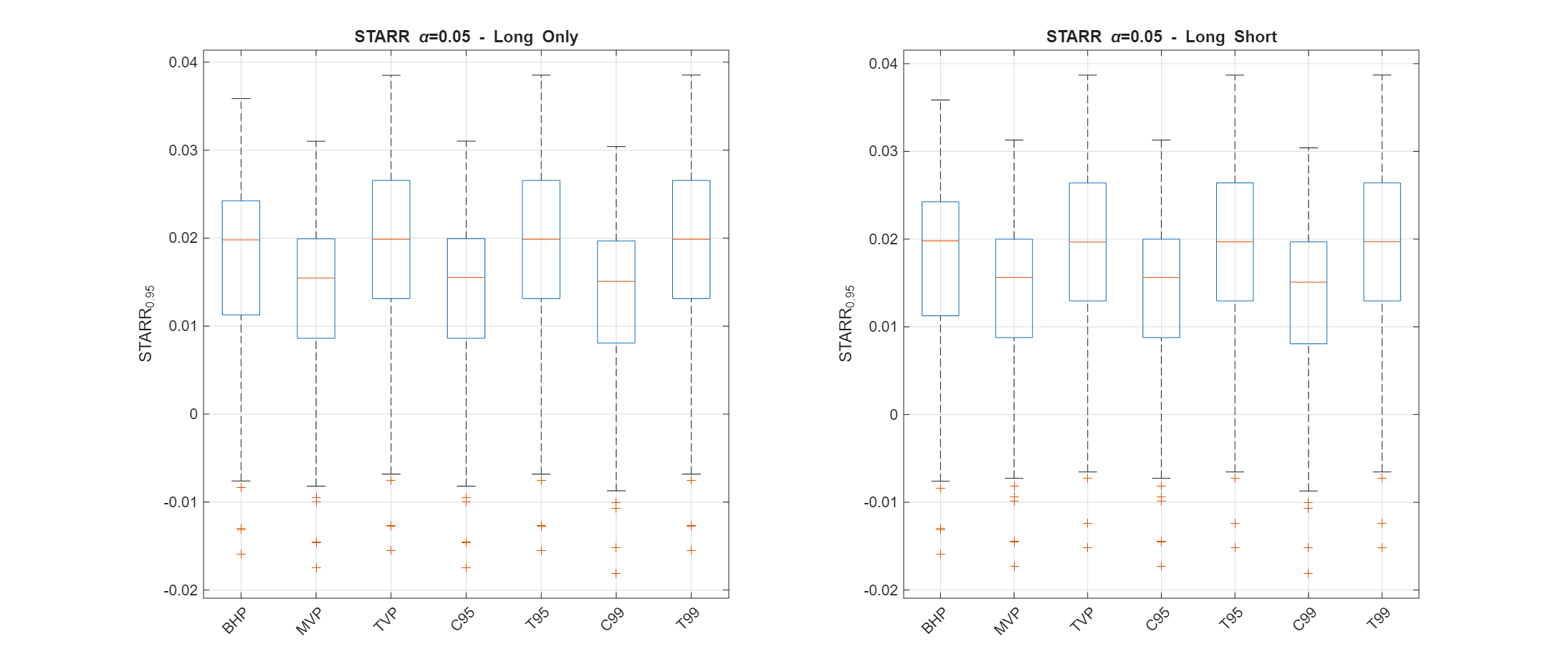}
\caption{STARR ratio distributions at the 95\% confidence level for the long-only and long--short portfolio strategies.}
\label{figSTARR95LOLS}
\end{figure}

\begin{figure}[h!]
\centering
\caption{STARR ratio distributions at the 99\% confidence level for the long-only and long--short portfolio strategies.}
\includegraphics[width=\linewidth]{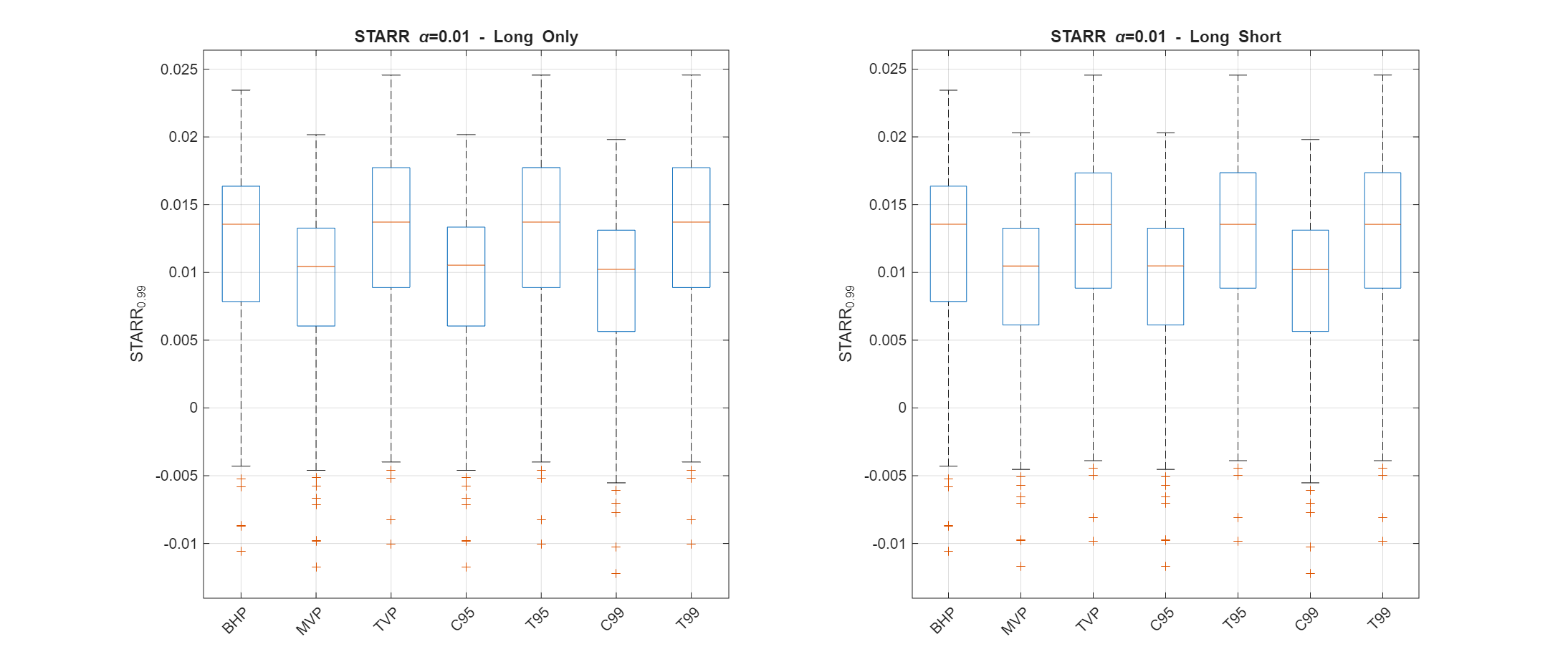}
\label{figSTARR99LOLS}
\end{figure}

\begin{figure}[h!]
\centering
\caption{Rachev ratio distributions at the 5\% tail level for the long-only and long--short portfolio strategies.}
\includegraphics[width=\linewidth]{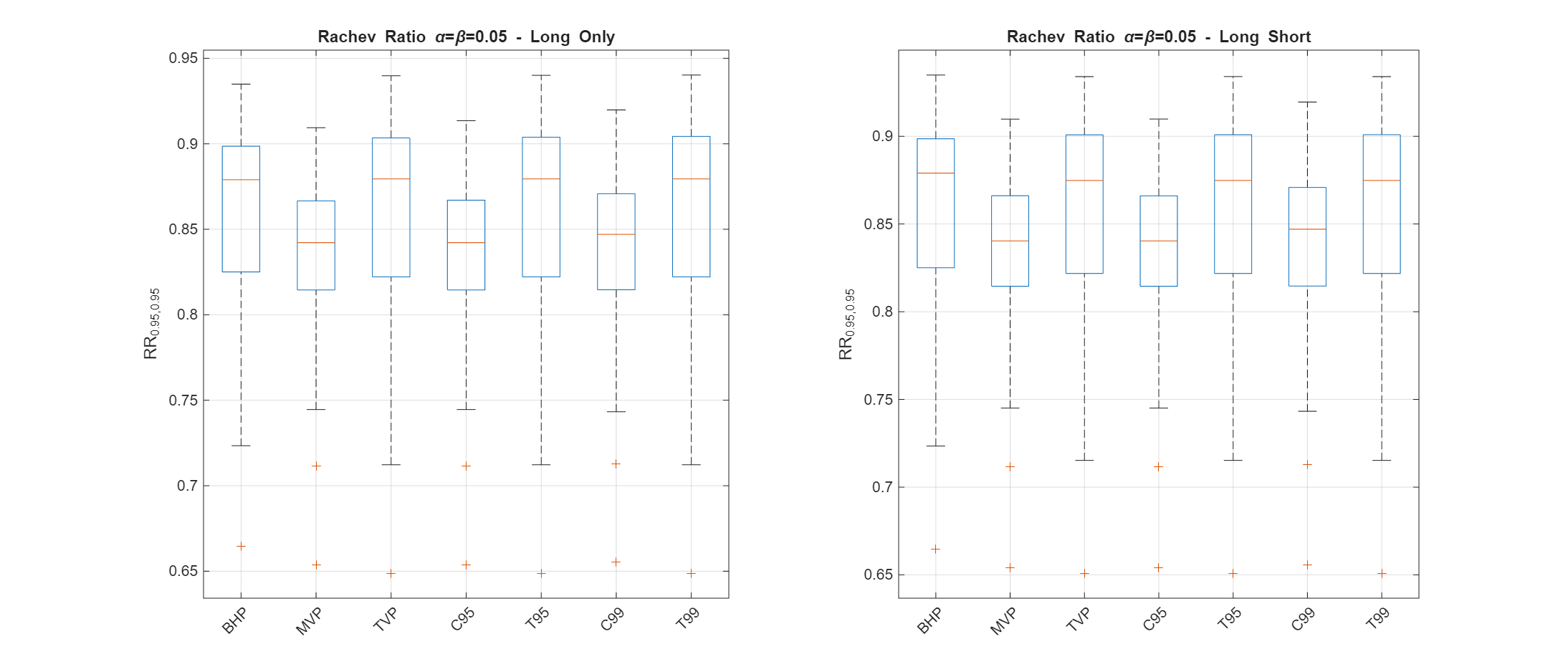}
\label{figRachev95LOLS}
\end{figure}

\begin{figure}[h!]
\centering
\includegraphics[width=\linewidth]{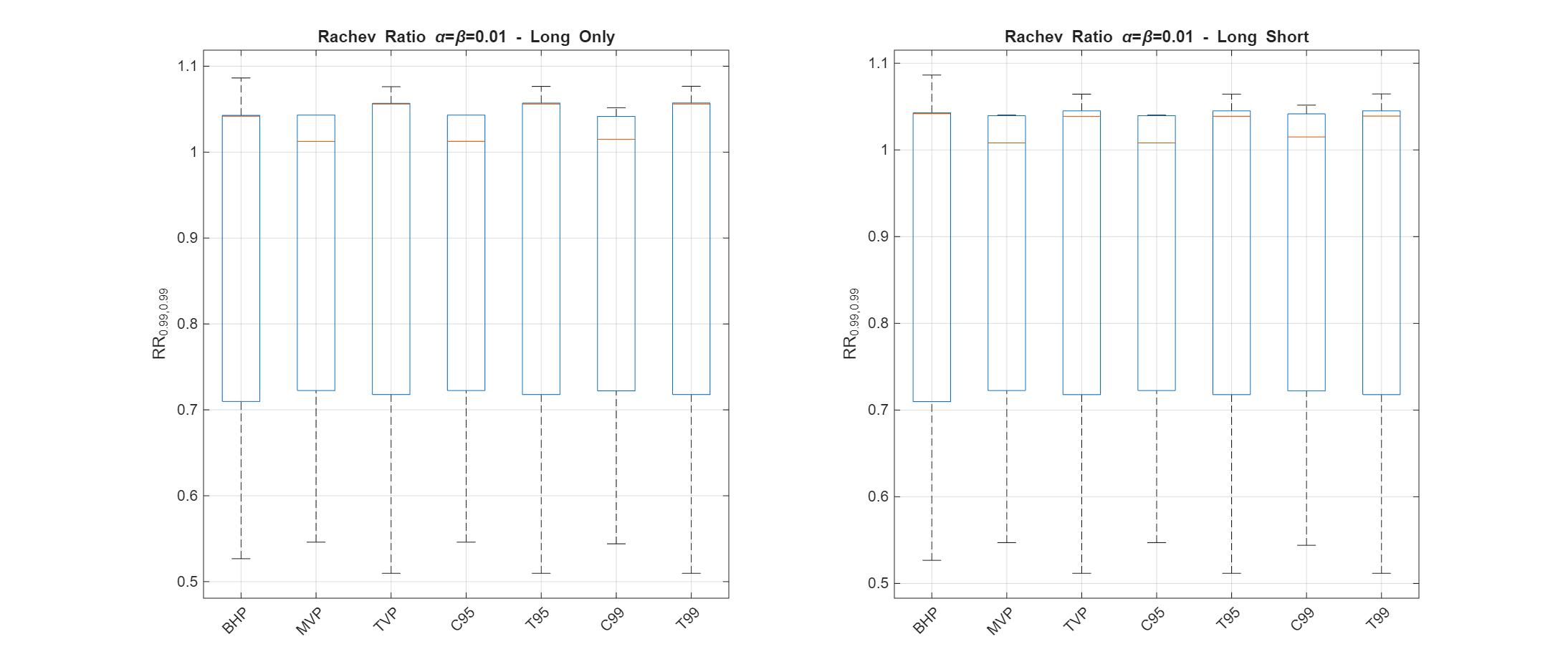}
\caption{Rachev ratio distributions at the 1\% tail level for the long-only and long--short portfolio strategies.}
\label{figRachev99LOLS}
\end{figure}

To evaluate whether this pattern persists after adjusting for risk, Figures~\ref{figSharpeLOLS}--\ref{figRachev99LOLS} report the Sharpe, Sortino, Calmar, STARR, and Rachev ratio distributions for the historical long-only and long--short portfolios. These measures compare the strategies across total volatility, downside variation, maximum drawdown, expected tail loss, and the balance between favorable upper-tail outcomes and unfavorable lower-tail outcomes \citep{SortinoVanDerMeer1991,Young1991,MartinRachevSiboulet2003,RachevEtAl2008}.

\FloatBarrier
\FloatBarrier

The risk-adjusted results reveal a consistent pattern. Tangency-type portfolios are generally competitive across volatility-based, downside-sensitive, drawdown-based, and tail-based performance measures. The buy-and-hold benchmark remains a meaningful comparator and is not uniformly dominated. Minimum-variance and CVaR-minimizing portfolios tend to produce more conservative return profiles, often sacrificing upside participation in exchange for stronger risk containment. The similarity between long-only and long--short distributions further indicates that allowing short positions does not automatically improve historical risk-adjusted performance under the bounds imposed in this study.
\begin{table}[h!]
\centering
\caption{Bootstrap Tests of Historical Sharpe-Ratio Differences}
\label{tabSharpeTestsHistorical}
\small
\begin{tabular}{lrr}
\toprule
Comparison & Sharpe Difference & Bootstrap (p)-value \\
\midrule
LO: MVP vs C95  & -0.00062 & 0.273 \\
LO: MVP vs C99  & -0.00058 & 0.734 \\
LO: TVP vs TC95 &  0.00000 & 0.967 \\
LO: TVP vs TC99 &  0.00012 & 0.293 \\
LS: MVP vs C95  & -0.00004 & 0.677 \\
LS: MVP vs C99  & -0.00036 & 0.864 \\
LS: TVP vs TC95 & -0.00006 & 0.079 \\
LS: TVP vs TC99 & -0.00003 & 0.716 \\
\bottomrule
\end{tabular}
\end{table}

To assess whether Sharpe-ratio differences are statistically meaningful, Table~\ref{tabSharpeTestsHistorical} reports bootstrap tests comparing mean--variance and CVaR-based historical portfolios. Because portfolio returns may exhibit non-normality, skewness, and heavy tails, bootstrap inference is preferred to standard asymptotic Sharpe-ratio tests \citep{EfronTibshirani1993,LedoitWolf2008}. Across both constraint regimes, the estimated differences are small and statistically insignificant at conventional levels. This does not imply that the strategies are economically equivalent; rather, it shows that variance-based performance measures alone do not fully capture differences in downside-risk and tail-risk behavior.

\subsection{Dynamic Portfolio Optimization Results}
\label{subsecDynamicPortfolioResults}

The dynamic allocation analysis examines how portfolio performance changes when weights are updated through a three-year rolling estimation window. At each rebalancing date, portfolio inputs are estimated using only information available up to that date, and the resulting weights are applied to subsequent realized returns. This design allows the allocation rule to respond to changing return, volatility, dependence, and downside-risk conditions through time.

\begin{figure}[h!]
\centering
\includegraphics[width=\linewidth]{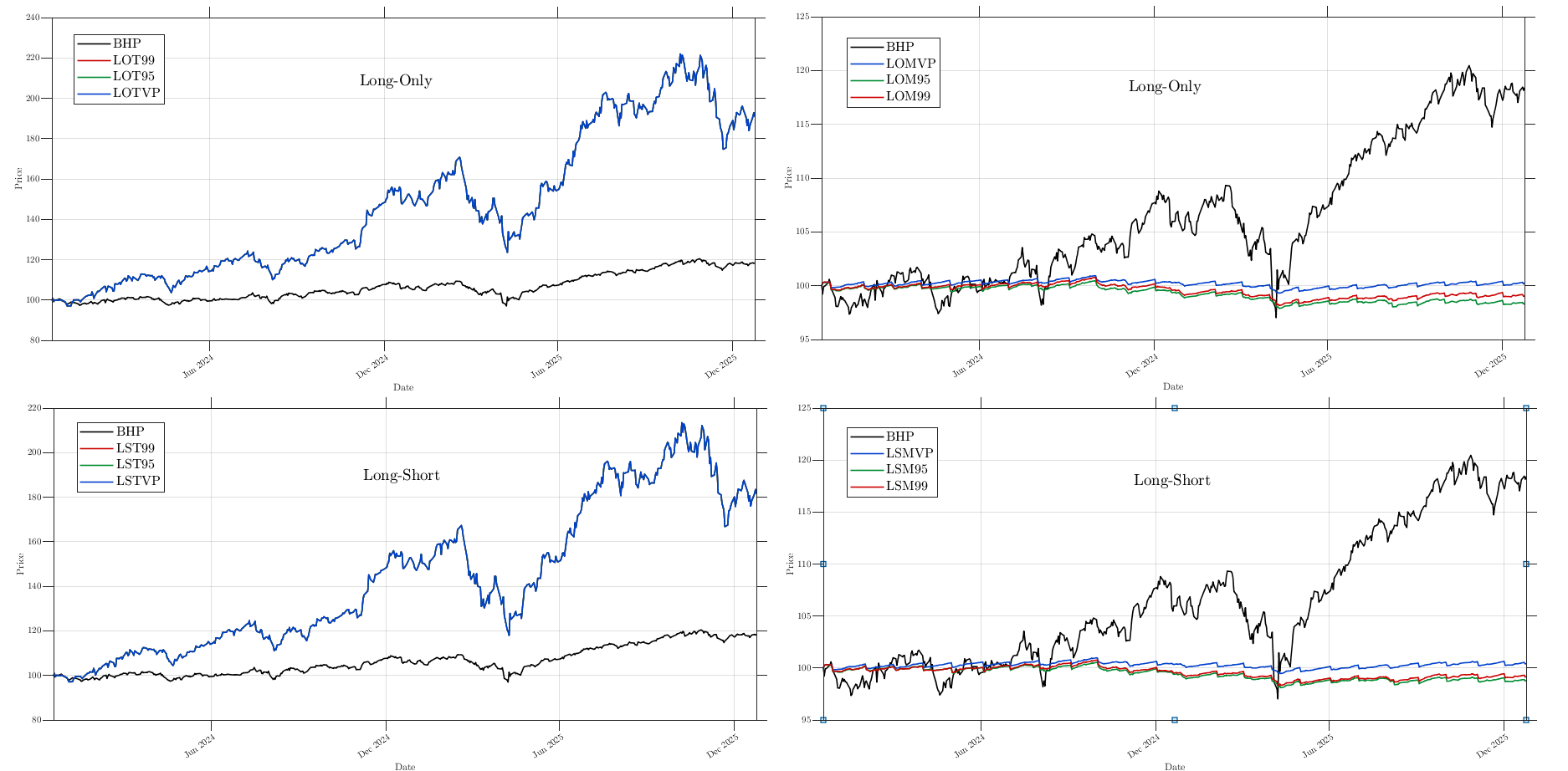}
\caption{Indexed cumulative values of dynamically optimized long-only and long--short portfolios, with all series normalized to 100 at the start of the dynamic evaluation period.}
\label{figDynamicIndexed}
\end{figure}

Figure~\ref{figDynamicIndexed} presents indexed cumulative values of the dynamically optimized portfolios under long-only and long--short constraints, benchmarked against the buy-and-hold portfolio. The dynamic tangency portfolio is the main source of cumulative outperformance. In both panels, it rises substantially above the benchmark and the more conservative optimized portfolios, indicating that the return-seeking dynamic allocation captures the largest upside potential over the sample. By contrast, the dynamic minimum-variance and CVaR-minimizing portfolios display flatter cumulative paths because they prioritize risk containment rather than wealth maximization.

\begin{figure}[h!]
\centering
\includegraphics[width=\textwidth]{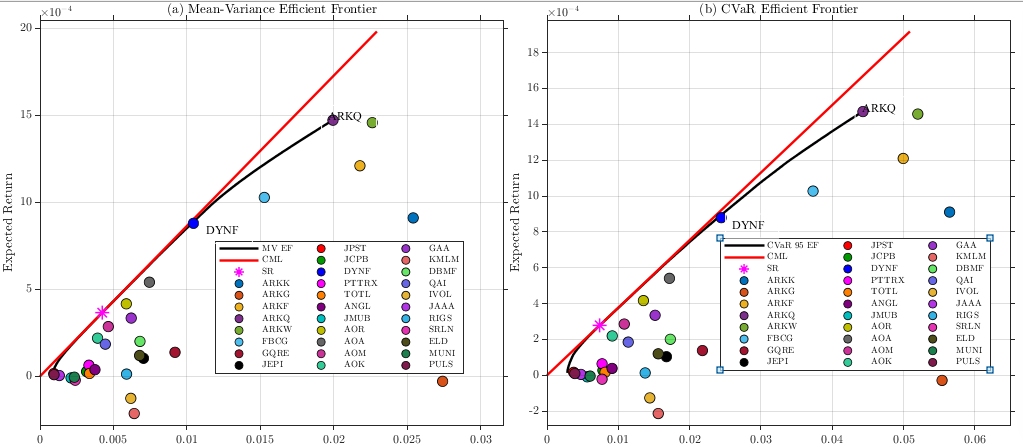}
\caption{Dynamic mean--variance and CVaR efficient frontiers with corresponding capital market lines and individual funds. The starred point denotes the Sharpe-ratio-maximizing portfolio in each case.}
\label{figDynamicFrontiers}
\end{figure}

Figure~\ref{figDynamicFrontiers} reports the dynamic mean--variance and CVaR efficient frontiers with corresponding capital market lines and individual funds. The frontiers show that diversified portfolio combinations continue to dominate many individual funds, confirming that portfolio construction remains valuable in the rolling-window setting. The difference between the mean--variance and CVaR panels again shows that portfolio rankings depend not only on expected return, but also on how risk is measured.

\begin{table}[h!]
\centering
\caption{Sharpe Ratios: Historical versus Dynamic Portfolios}
\label{tabDynSharpe}
\small
\begin{tabular}{lcccc}
\toprule
Strategy & Hist LO & Dyn LO & Hist LS & Dyn LS \\
\midrule
MVP  & 1.1971 & 0.7007 & 1.4365 & 1.2818 \\
TVP  & 1.1468 & 0.7747 & 3.1441 & 0.8379 \\
C95  & 1.2958 & 1.0660 & 1.7773 & 1.3818 \\
TC95 & 1.1022 & 0.4764 & 3.1584 & 1.7172 \\
C99  & 1.3635 & 0.7529 & 1.6713 & 1.5489 \\
TC99 & 1.0544 & 0.3605 & 3.0517 & 1.8306 \\
\bottomrule
\end{tabular}
\end{table}

\begin{table}[h!]
\centering
\caption{Calmar Ratios: Historical versus Dynamic Portfolios}
\label{tabDynCalmar}
\small
\begin{tabular}{lcccc}
\toprule
Strategy & Hist LO & Dyn LO & Hist LS & Dyn LS \\
\midrule
MVP  & 2.3156 & 1.2340 & 2.3182 & 2.1386 \\
TVP  & 2.0159 & 1.1177 & 5.6147 & 1.5858 \\
C95  & 2.4719 & 1.8153 & 2.9027 & 2.1471 \\
TC95 & 2.0490 & 0.7899 & 5.3253 & 3.1222 \\
C99  & 2.6480 & 1.2075 & 2.5774 & 2.1081 \\
TC99 & 1.8370 & 0.6490 & 7.5406 & 2.4624 \\
\bottomrule
\end{tabular}
\end{table}

\begin{table}[h!]
\centering
\caption{STARR Ratios: Historical versus Dynamic Portfolios}
\label{tabDynSTARR}
\small
\begin{tabular}{lcccc}
\toprule
Strategy & Hist LO & Dyn LO & Hist LS & Dyn LS \\
\midrule
MVP  & 11.0830 & 7.9086 & 13.1997 & 11.1278 \\
TVP  & 10.3560 & 6.9143 & 27.5372 & 8.2569 \\
C95  & 11.8507 & 11.0352 & 16.9871 & 11.7725 \\
TC95 & 10.0362 & 5.6006 & 29.6151 & 15.7983 \\
C99  & 12.4632 & 8.6207 & 15.9424 & 13.0333 \\
TC99 & 9.5653 & 4.9927 & 30.5929 & 16.5621 \\
\bottomrule
\end{tabular}
\end{table}

\begin{table}[h!]
\centering
\caption{Rachev Ratios: Historical versus Dynamic Portfolios}
\label{tabDynRachev}
\small
\begin{tabular}{lcccc}
\toprule
Strategy & Hist LO & Dyn LO & Hist LS & Dyn LS \\
\midrule
MVP  & 0.9939 & 0.9974 & 1.0170 & 0.9524 \\
TVP  & 1.0376 & 1.0648 & 1.3094 & 1.1443 \\
C95  & 1.0211 & 1.0630 & 1.1642 & 0.9572 \\
TC95 & 1.0508 & 1.0249 & 1.4200 & 1.1893 \\
C99  & 1.0161 & 1.0582 & 1.1740 & 0.9898 \\
TC99 & 1.0692 & 1.0733 & 1.5550 & 1.2852 \\
\bottomrule
\end{tabular}
\end{table}
Tables~\ref{tabDynSharpe}--\ref{tabDynRachev} compare historical and dynamic portfolios using Sharpe, Calmar, STARR, and Rachev ratios. In the long-only setting, the dynamic C95 portfolio is the most consistently attractive dynamic specification across Sharpe, Calmar, and STARR criteria. In the long--short setting, TC95 and TC99 dominate most other dynamic strategies across several risk-adjusted measures. These findings show that dynamic optimization does not improve all strategies uniformly; its value depends on the interaction between the optimization objective and the admissible constraint set.

\begin{figure}[h!]
\centering
\includegraphics[width=.9\textwidth]{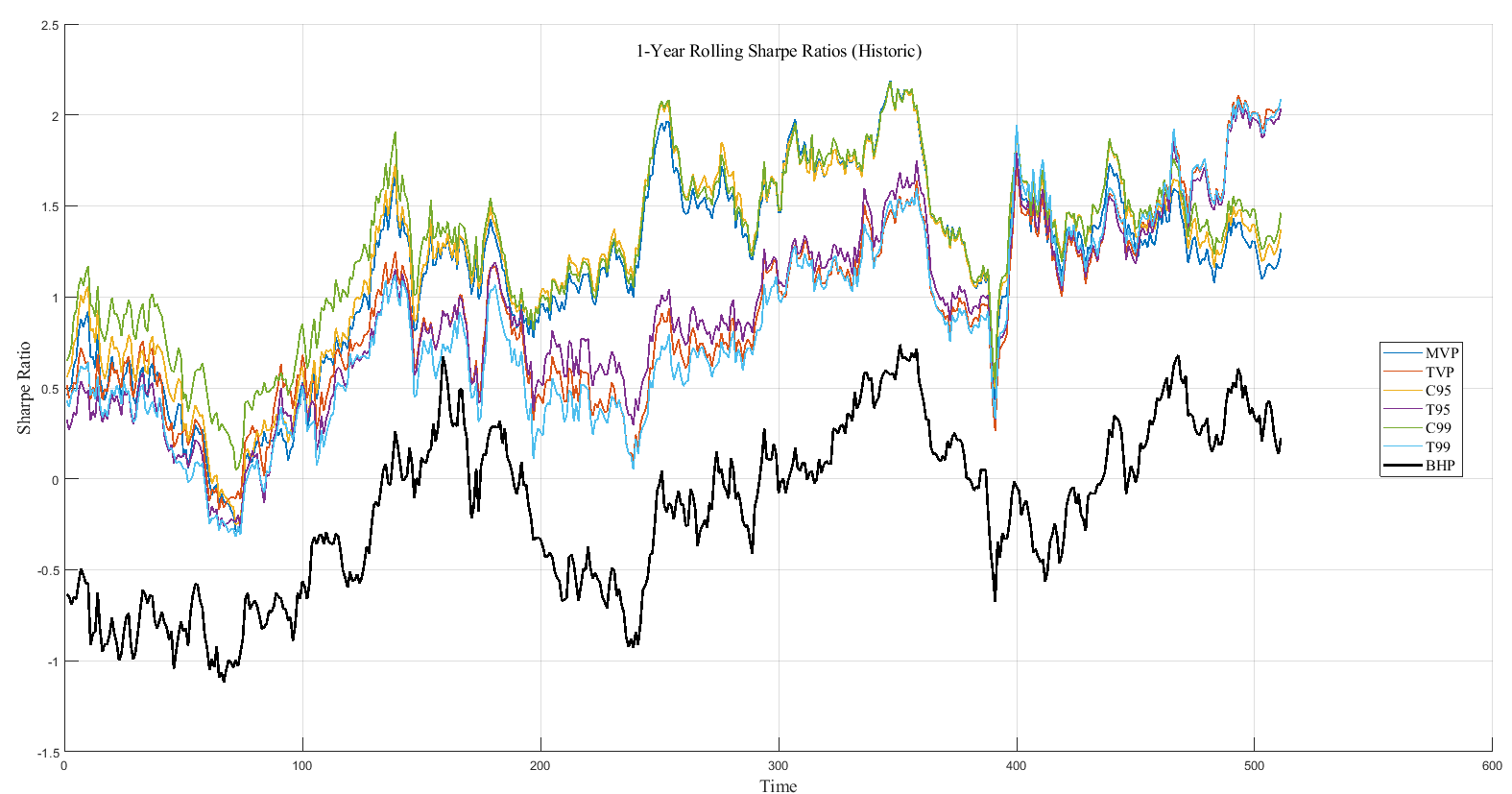}
\caption{One-year rolling Sharpe ratios for the historical portfolios.}
\label{figRollingSharpeHistorical}
\end{figure}

\begin{figure}[h!]
\centering
\includegraphics[width=.9\textwidth]{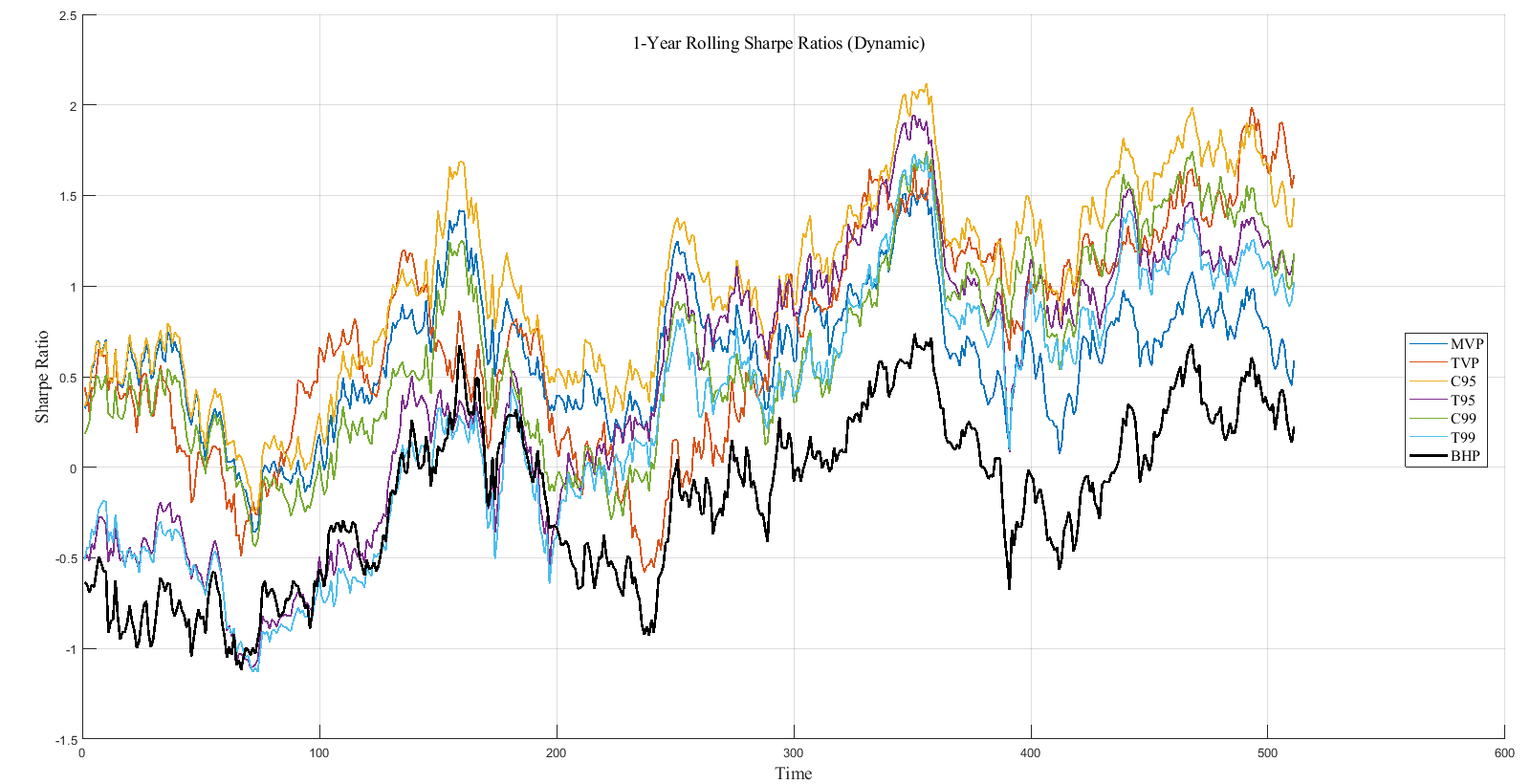}
\caption{One-year rolling Sharpe ratios for the dynamic portfolios.}
\label{figRollingSharpeDynamic}
\end{figure}

Figures~\ref{figRollingSharpeHistorical} and~\ref{figRollingSharpeDynamic} provide a time-varying view of performance by plotting one-year rolling Sharpe ratios. The historical optimized portfolios generally remain above the buy-and-hold benchmark over much of the rolling evaluation period, while the dynamic portfolios display greater dispersion. Dynamic C95 provides the most consistent long-only improvement, whereas dynamic tangency-type strategies offer stronger upside potential but with greater variation through time.

Because dynamic strategies require rebalancing, transaction costs may materially affect implementable performance. We compute turnover as
\begin{equation}
\mathrm{Turnover}_{t}
=\frac{1}{2}
\sum_{i=1}^{N}
\left|w_{i,t}-w_{i,t-1}\right|,
\label{eqTurnover}
\end{equation}
and net portfolio returns as
\begin{equation}
R^{\mathrm{net}}_{p,t}
=R^{\mathrm{gross}}_{p,t}
\frac{c}{10000}\mathrm{Turnover}_{t},
\label{eqNetReturn}
\end{equation}
where (c) is the transaction-cost rate in basis points.

Table~\ref{tabDynamicTCRobustness} reports transaction-cost-adjusted results using a 50 basis-point cost assumption. Dynamic rebalancing can materially reduce realized performance when turnover is high. The effect is especially pronounced for dynamic tangency-type portfolios, which retain positive net terminal wealth gains but experience substantial wealth drag. MVP and CVaR-minimizing dynamic portfolios perform poorly after transaction costs because their gross returns are already weak and turnover further reduces net performance.

\begin{table}[h!]
\centering
\caption{Dynamic Portfolio Transaction-Cost Robustness at 50 Basis Points}
\label{tabDynamicTCRobustness}
\scriptsize
\begin{tabular}{llrrrrr}
\toprule
Constraint & Strategy & Ann. Turnover (\%) & Gross TW & Net TW & Wealth Drag & Net Sharpe \\
\midrule
BHP & BHP & 0.0 & 118.42 & 118.42 & 0.00 & 0.507 \\
LO & MVP & 1733.0 & 99.95 & 84.13 & 15.82 & -8.686 \\
LO & TVP/TC95/TC99 & 3873.1 & 177.07 & 120.50 & 56.57 & 0.331 \\
LO & C95 & 2701.6 & 98.08 & 74.97 & 23.11 & -11.027 \\
LO & C99 & 2993.7 & 98.76 & 73.33 & 25.43 & -12.219 \\
LS & MVP & 1736.1 & 100.17 & 84.29 & 15.88 & -8.613 \\
LS & TVP/TC95/TC99 & 3842.9 & 168.30 & 114.87 & 53.43 & 0.239 \\
LS & C95 & 2706.1 & 98.53 & 75.28 & 23.25 & -11.008 \\
LS & C99 & 3001.0 & 98.90 & 73.38 & 25.52 & -12.319 \\
\bottomrule
\end{tabular}

\vspace{0.4em}
\parbox{0.94\textwidth}{\scriptsize \textit{Notes:} TW denotes terminal wealth from an initial value of 100. Net TW is computed after applying a 50 basis-point transaction cost per unit of turnover. TVP, TC95, and TC99 have identical dynamic transaction-cost results in the workbook and are therefore combined to save space.}
\end{table}

Table~\ref{tabDynamicSharpeTests} presents bootstrap Sharpe-ratio difference tests for the dynamic allocation strategies. The null and alternative hypotheses are
\begin{equation}
H_0: SR_i-SR_j=0,
\qquad
H_A: SR_i-SR_j\neq0.
\label{eqSharpeHypotheses}
\end{equation}
The results indicate that the observed differences between mean--variance and CVaR-based strategies are not statistically significant at conventional levels. Thus, when evaluated only through the traditional mean-return-to-volatility trade-off, the competing dynamic optimization approaches generate statistically comparable risk-adjusted performance.

\begin{table}[htbp]
\centering
\caption{Bootstrap Tests of Dynamic Portfolio Sharpe-Ratio Differences}
\label{tabDynamicSharpeTests}
\small
\begin{tabular}{lrr}
\toprule
Comparison & Sharpe Difference & Bootstrap (p)-value \\
\midrule
LO: MVP vs C95  & 0.0374 & 0.520 \\
LO: MVP vs C99  & 0.0239 & 0.505 \\
LO: TVP vs TC95 & 0.0000 & 1.000 \\
LO: TVP vs TC99 & 0.0000 & 1.000 \\
LS: MVP vs C95  & 0.0334 & 0.512 \\
LS: MVP vs C99  & 0.0260 & 0.514 \\
LS: TVP vs TC95 & 0.0000 & 1.000 \\
LS: TVP vs TC99 & 0.0000 & 1.000 \\
\bottomrule
\end{tabular}
\end{table}

The portfolio results can be interpreted through the interaction of fund heterogeneity, optimization criteria, and portfolio constraints. Tangency-type portfolios emphasize reward-to-risk maximization and can generate stronger upside-oriented performance, but may also carry greater vulnerability to adverse tail events. Minimum-variance and CVaR-minimizing portfolios are more defensive, sacrificing upside participation for volatility reduction or downside-loss control. Long-only constraints are more consistent with conventional investment mandates, whereas long--short constraints expand the feasible set but may increase estimation sensitivity, turnover, financing costs, and implementation complexity. Overall, actively managed ETFs should be evaluated as interacting portfolio components rather than as isolated funds, and no single optimization rule dominates across all dimensions.

\begin{table}[htbp]
\centering
\caption{Economic Interpretation of Portfolio Allocation Mechanisms}
\label{tabEconomicInterpretation}
\scriptsize
\begin{tabular}{lll}
\toprule
Portfolio Feature & Economic Mechanism & Investor-Relevant Interpretation \\
\midrule
Tangency portfolios & Reward-to-risk maximization & Higher upside potential, but possibly larger tail exposure \\
Minimum-variance portfolios & Volatility reduction & Defensive allocation with lower total-risk exposure \\
CVaR portfolios & Downside-loss control & Tail-risk mitigation and drawdown protection \\
Long-only constraints & Restricted feasible set & More implementable and mandate-consistent allocations \\
Long--short constraints & Expanded feasible set & Greater flexibility but higher implementation complexity \\
Dynamic allocation & Rolling information updates & Adaptive exposure to changing market conditions \\
\bottomrule
\end{tabular}
\end{table}

\section{Downside-Risk and Tail-Risk Analytics}
\label{secTailRisk}

This section examines the downside-risk and lower-tail behavior of the historical and dynamic portfolios. The analysis combines empirical Value at Risk (VaR), Expected Shortfall (ES)/Tail Conditional Expectation (TCE), maximum drawdown, left-tail Hill estimators \citep{Hill1975}, and POT--GPD diagnostics within the broader extreme-value framework for heavy-tailed financial losses \citep{EmbrechtsKluppelbergMikosch1997,McNeilFreyEmbrechts2015}.

\subsection{Empirical Downside-Risk Measures}
\label{subsecEmpiricalTailRisk}

Let \(R_{p,t}\) denote the daily portfolio return and define the loss variable as
\begin{equation}
L_{p,t}=-R_{p,t}.
\label{eqPortfolioLoss}
\end{equation}
For confidence level \(\alpha\), empirical Value-at-Risk and Expected Shortfall are computed as
\begin{equation}
\operatorname{VaR}_{\alpha}(L)=Q_{\alpha}(L),
\qquad
\operatorname{ES}_{\alpha}(L)
=\operatorname{TCE}_{\alpha}(L)=
\operatorname{E}\left[
L \mid L \geq \operatorname{VaR}_{\alpha}(L)
\right].
\label{eqVaRESTCE}
\end{equation}
\(Q_{\alpha}(L)\) denotes the \(\alpha\)-quantile of the loss distribution \(L\).\\

In this finite-sample implementation, ES and TCE are equivalent and measure the average severity of losses beyond the VaR threshold.

Tables~\ref{tabHistoricalTailRisk} and~\ref{tabDynamicTailRisk} report empirical VaR, ES/TCE, and maximum drawdown values for the historical and dynamic portfolios. These measures show that portfolios with similar conventional risk-adjusted performance can differ substantially in downside-loss profiles. In particular, dynamic tangency-type portfolios display materially larger tail losses and drawdowns than the corresponding minimum-risk dynamic portfolios, despite their stronger upside-oriented performance.

\begin{table}[htbp]
\centering
\caption{Historical Portfolio Tail-Risk Measures}
\label{tabHistoricalTailRisk}
\scriptsize
\begin{tabular}{llrrrrr}
\toprule
Constraint & Portfolio & VaR 95\% & ES/TCE 95\% & VaR 99\% & ES/TCE 99\% & Max DD \\
\midrule
LO & MVP  & 0.812 & 1.155 & 1.435 & 1.741 & 9.489 \\
LO & TVP  & 0.905 & 1.325 & 1.638 & 2.024 & 10.918 \\
LO & C95  & 0.812 & 1.155 & 1.435 & 1.741 & 9.489 \\
LO & TC95 & 0.905 & 1.325 & 1.636 & 2.024 & 10.918 \\
LO & C99  & 0.826 & 1.166 & 1.459 & 1.748 & 9.666 \\
LO & TC99 & 0.905 & 1.325 & 1.631 & 2.023 & 10.918 \\
LS & MVP  & 0.806 & 1.153 & 1.435 & 1.737 & 9.431 \\
LS & TVP  & 0.901 & 1.302 & 1.573 & 1.990 & 10.719 \\
LS & C95  & 0.806 & 1.153 & 1.435 & 1.737 & 9.431 \\
LS & TC95 & 0.901 & 1.302 & 1.573 & 1.990 & 10.719 \\
LS & C99  & 0.826 & 1.166 & 1.459 & 1.748 & 9.666 \\
LS & TC99 & 0.901 & 1.303 & 1.573 & 1.990 & 10.719 \\
\bottomrule
\end{tabular}
\end{table}

\begin{table}[htbp]
\centering
\caption{Dynamic Portfolio Tail-Risk Measures}
\label{tabDynamicTailRisk}
\scriptsize
\begin{tabular}{llrrrrr}
\toprule
Constraint & Portfolio & VaR 95\% & ES/TCE 95\% & VaR 99\% & ES/TCE 99\% & Max DD \\
\midrule
LO & MVP  & 0.178 & 0.358 & 0.412 & 0.423 & 1.615 \\
LO & TVP  & 2.563 & 3.798 & 4.801 & 5.295 & 27.620 \\
LO & C95  & 0.284 & 0.375 & 0.414 & 0.440 & 2.554 \\
LO & TC95 & 2.563 & 3.798 & 4.801 & 5.295 & 27.620 \\
LO & C99  & 0.199 & 0.363 & 0.401 & 0.434 & 2.593 \\
LO & TC99 & 2.563 & 3.798 & 4.801 & 5.295 & 27.620 \\
LS & MVP  & 0.175 & 0.358 & 0.408 & 0.421 & 1.501 \\
LS & TVP  & 2.562 & 3.913 & 4.864 & 5.569 & 29.441 \\
LS & C95  & 0.269 & 0.370 & 0.404 & 0.428 & 2.422 \\
LS & TC95 & 2.562 & 3.913 & 4.864 & 5.569 & 29.441 \\
LS & C99  & 0.210 & 0.361 & 0.396 & 0.423 & 2.417 \\
LS & TC99 & 2.562 & 3.913 & 4.864 & 5.569 & 29.441 \\
\bottomrule
\end{tabular}
\end{table}

\subsection{Left-Tail Hill Diagnostics}
\label{subsecHillDiagnostics}

To examine extreme downside behavior, negative portfolio returns are transformed into positive losses so that the left tail of returns can be studied as the upper tail of the loss distribution. The Hill estimator is then computed across a range of upper-order statistics (k). Smaller tail-index estimates correspond to heavier tails and therefore greater exposure to extreme downside losses. Because the estimator is sensitive to the threshold choice, the results are interpreted diagnostically rather than as exact tail-index estimates.

\begin{figure}[htbp]
\centering
\includegraphics[width=.49\textwidth]{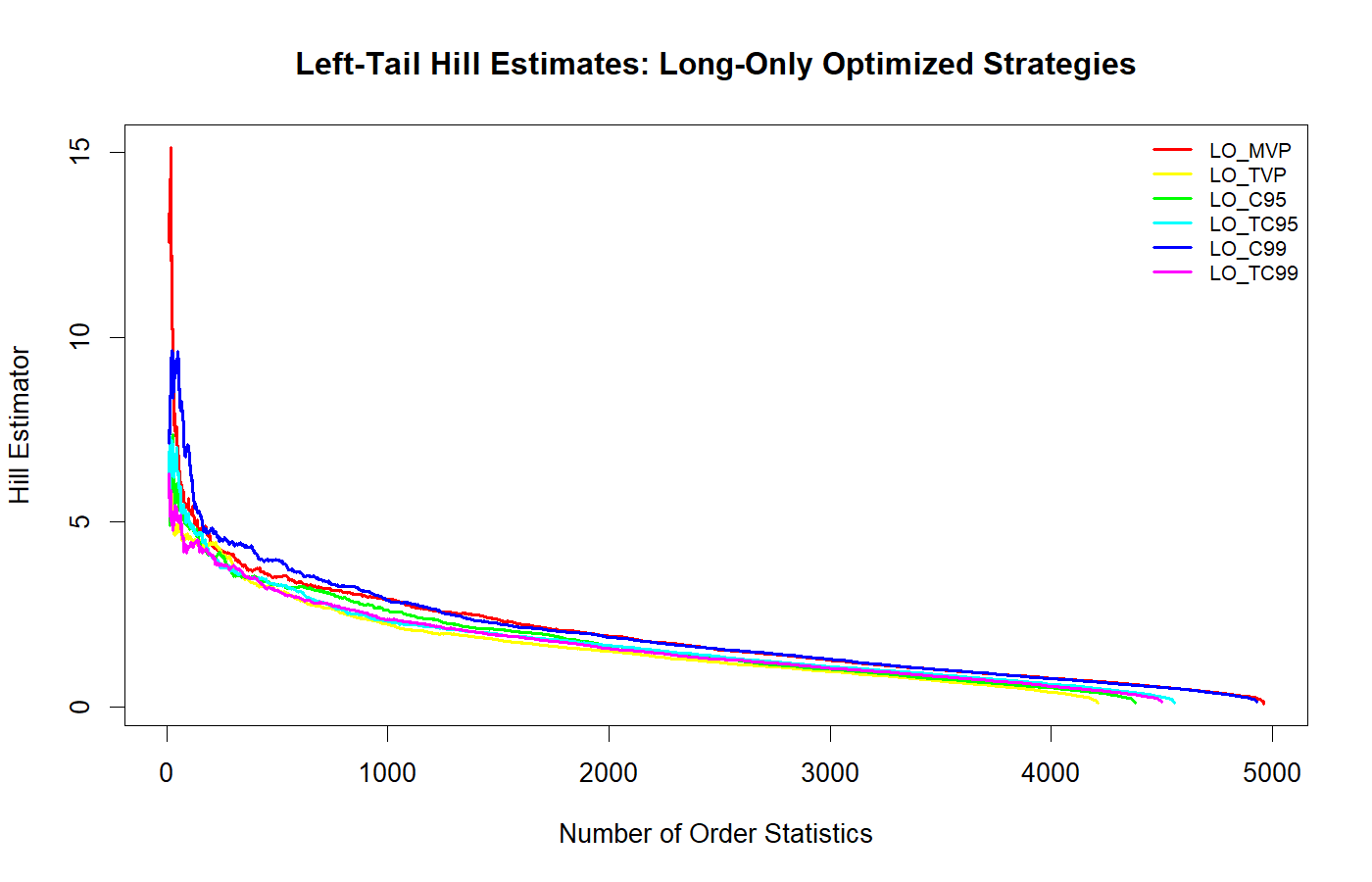}
\includegraphics[width=.49\textwidth]{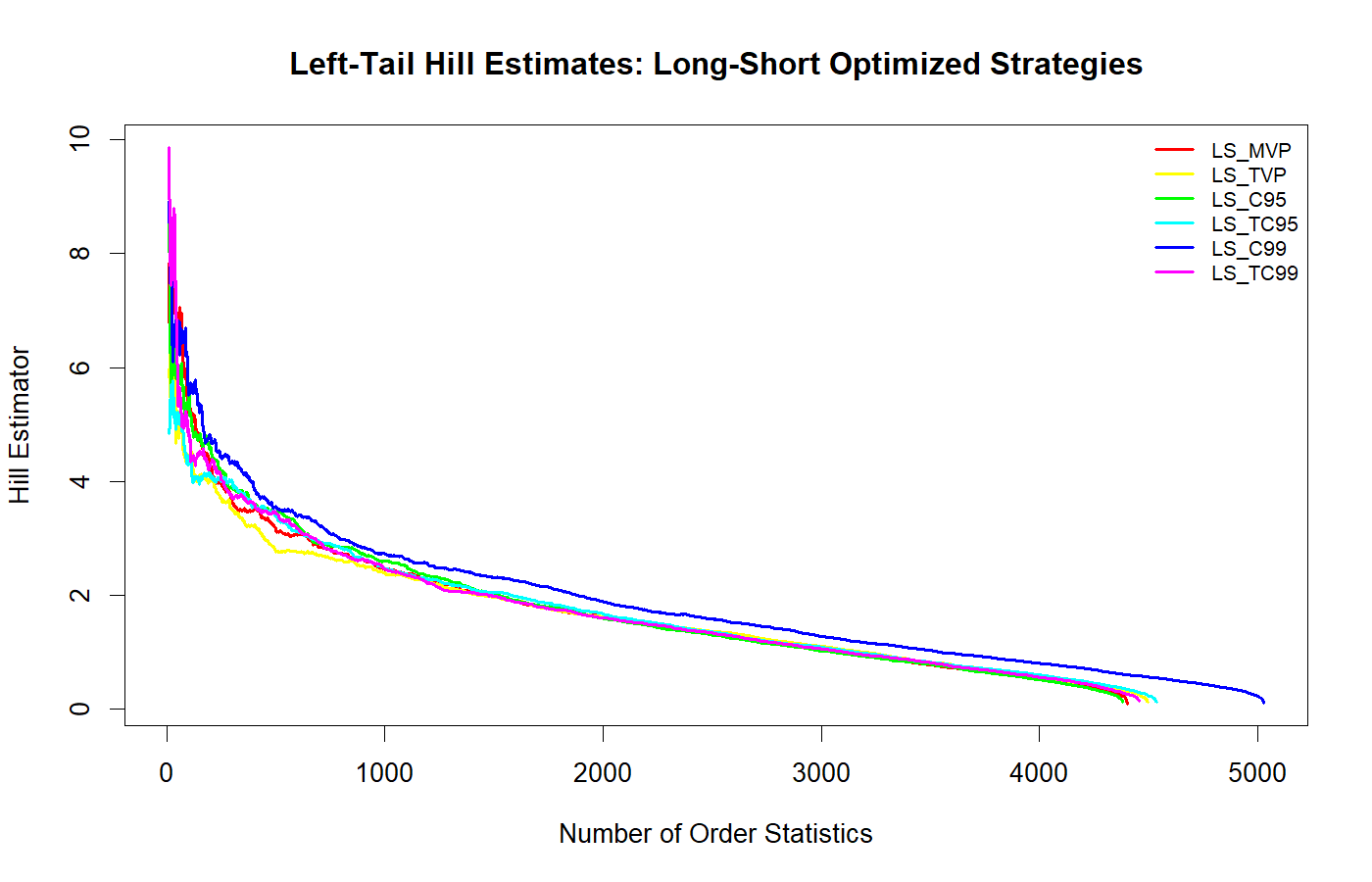}
\caption{Left-tail Hill estimators for historical long-only and long--short optimized strategies. Panel (a) reports long-only strategies, while Panel (b) reports long--short strategies.}
\label{figHillHistoricalCombined}
\end{figure}

Figure~\ref{figHillHistoricalCombined} reports the left-tail Hill estimates for the full set of historical optimized strategies under long-only and long--short constraints. The curves are more unstable at small threshold choices and become smoother as (k) increases, which is consistent with the usual bias--variance trade-off in tail-index estimation. The historical long-only strategies are relatively close over much of the threshold range, while the long--short strategies show somewhat greater separation. This indicates that shorting flexibility can change the behavior of extreme downside losses even when cumulative performance differences are modest.

\begin{figure}[h!]
\centering
\includegraphics[width=.8\textwidth]{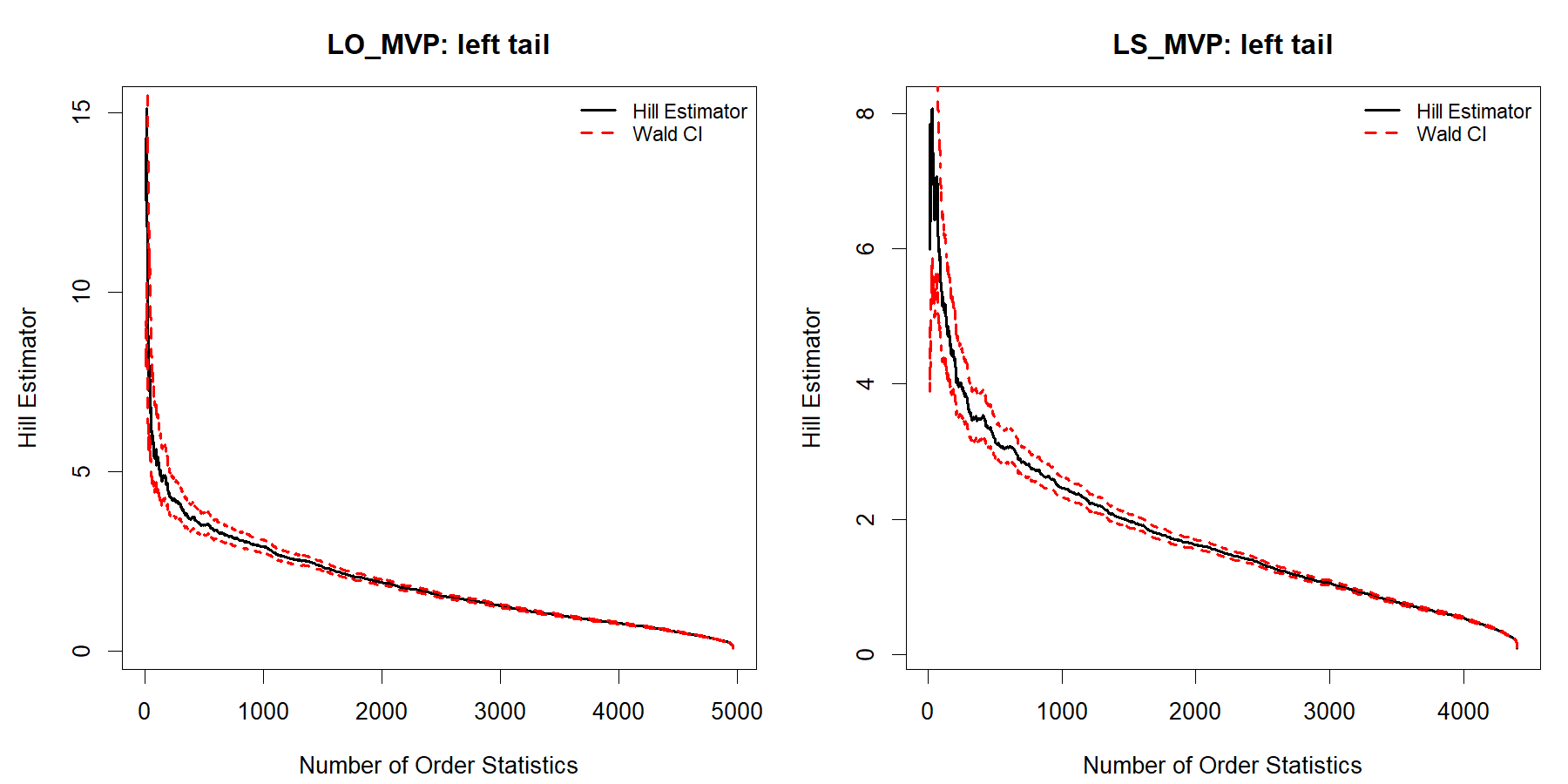}
\includegraphics[width=.8\textwidth]{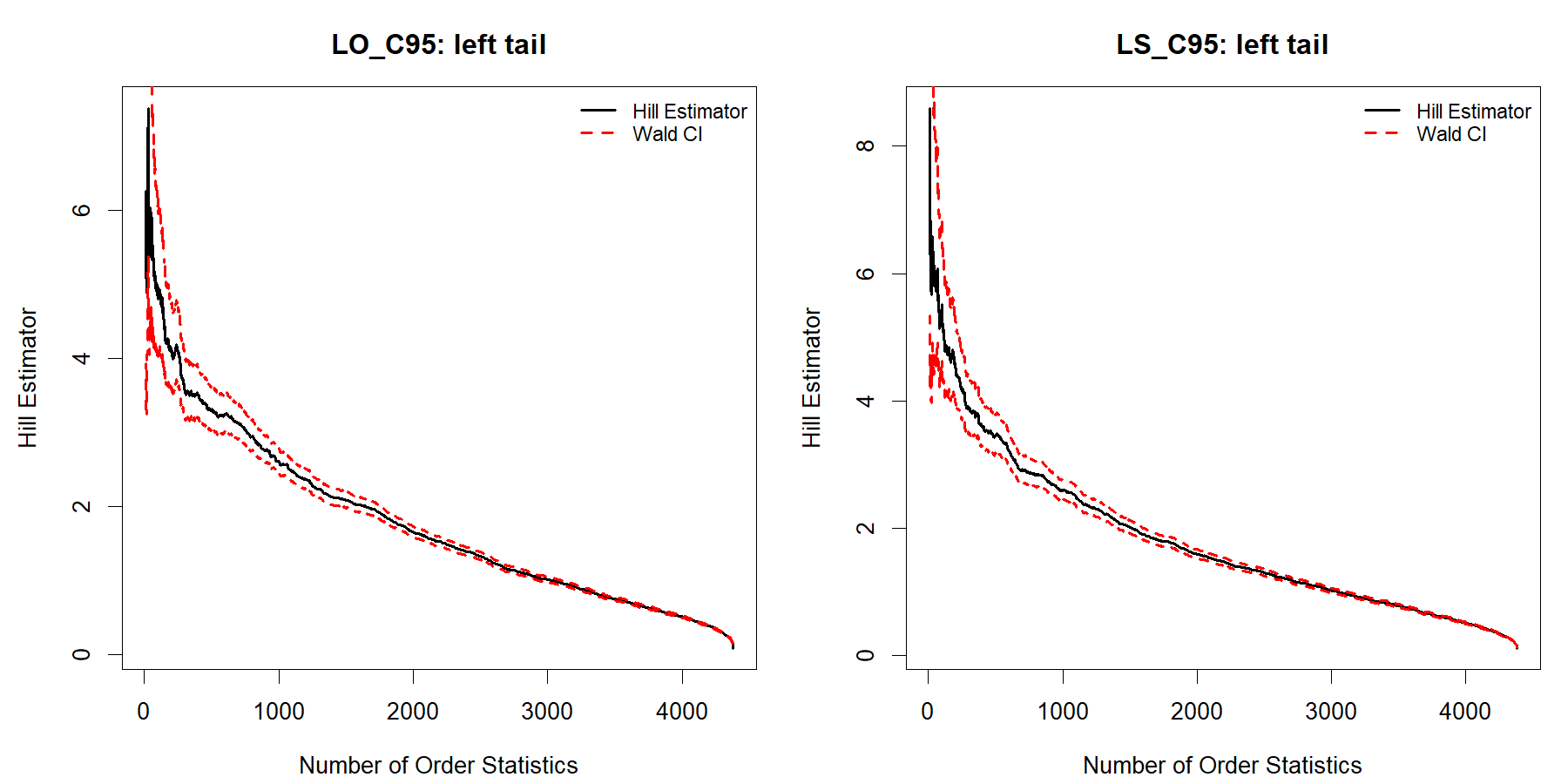}
\includegraphics[width=.8\textwidth]{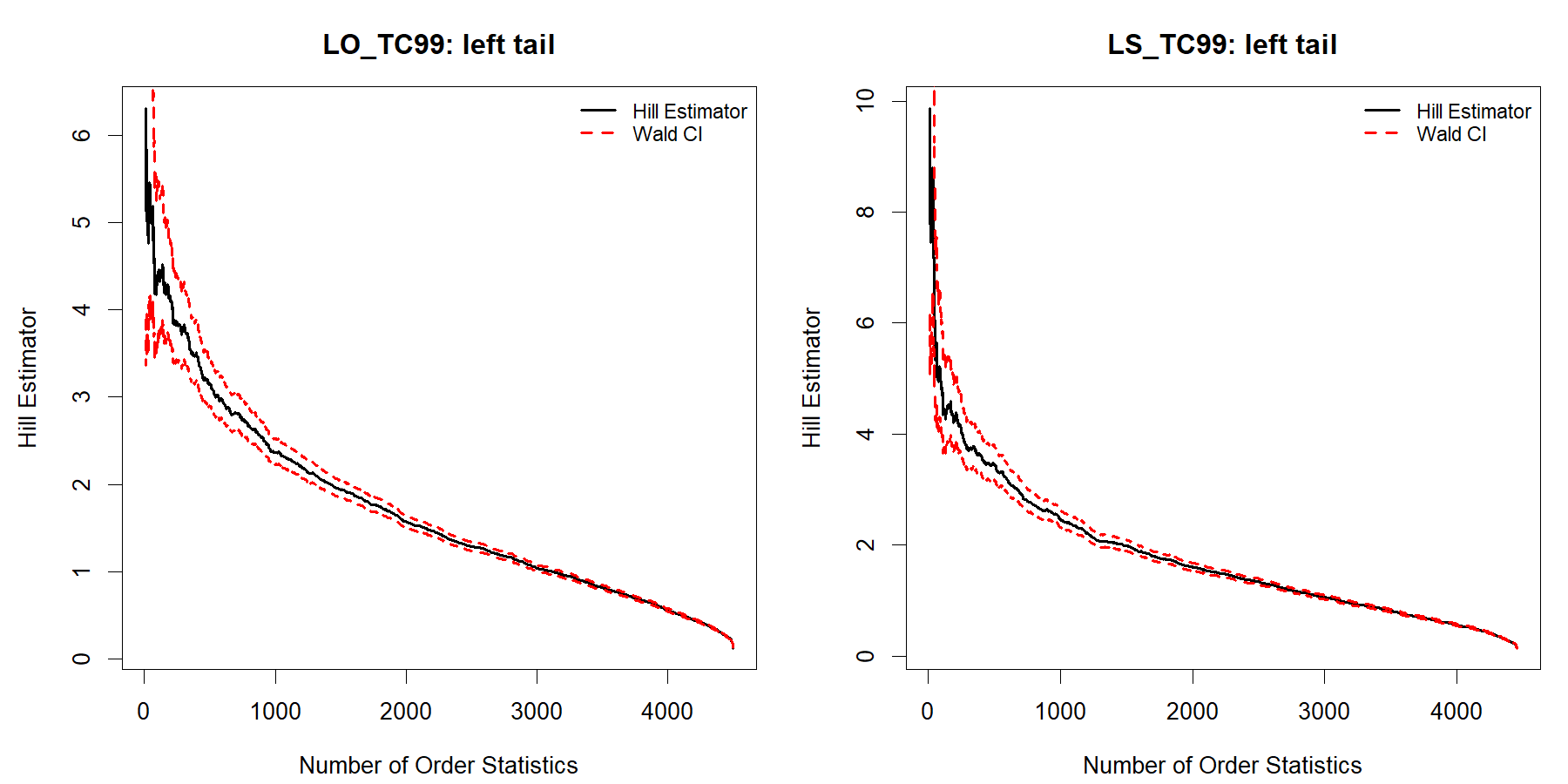}
\caption{Left-tail Hill comparisons for representative historical strategies. Panel (a) compares MVP, Panel (b) compares C95, and Panel (c) compares TC99.}
\label{figHillHistoricalSelected}
\end{figure}

Figure~\ref{figHillHistoricalSelected} compares representative historical strategy pairs: MVP, C95, and TC99. The long-only and long--short MVP curves are broadly similar for moderate and large threshold values but separate more visibly in the most extreme-loss region. The C95 curves remain closer across thresholds, suggesting that the 95\% CVaR objective imposes similar downside-tail discipline under both constraint regimes. The TC99 comparison shows more persistent separation, indicating that far-tail, return-seeking optimization combined with shorting flexibility can materially change downside-tail exposure.

\begin{figure}[htbp]
\centering
\includegraphics[width=.49\textwidth]{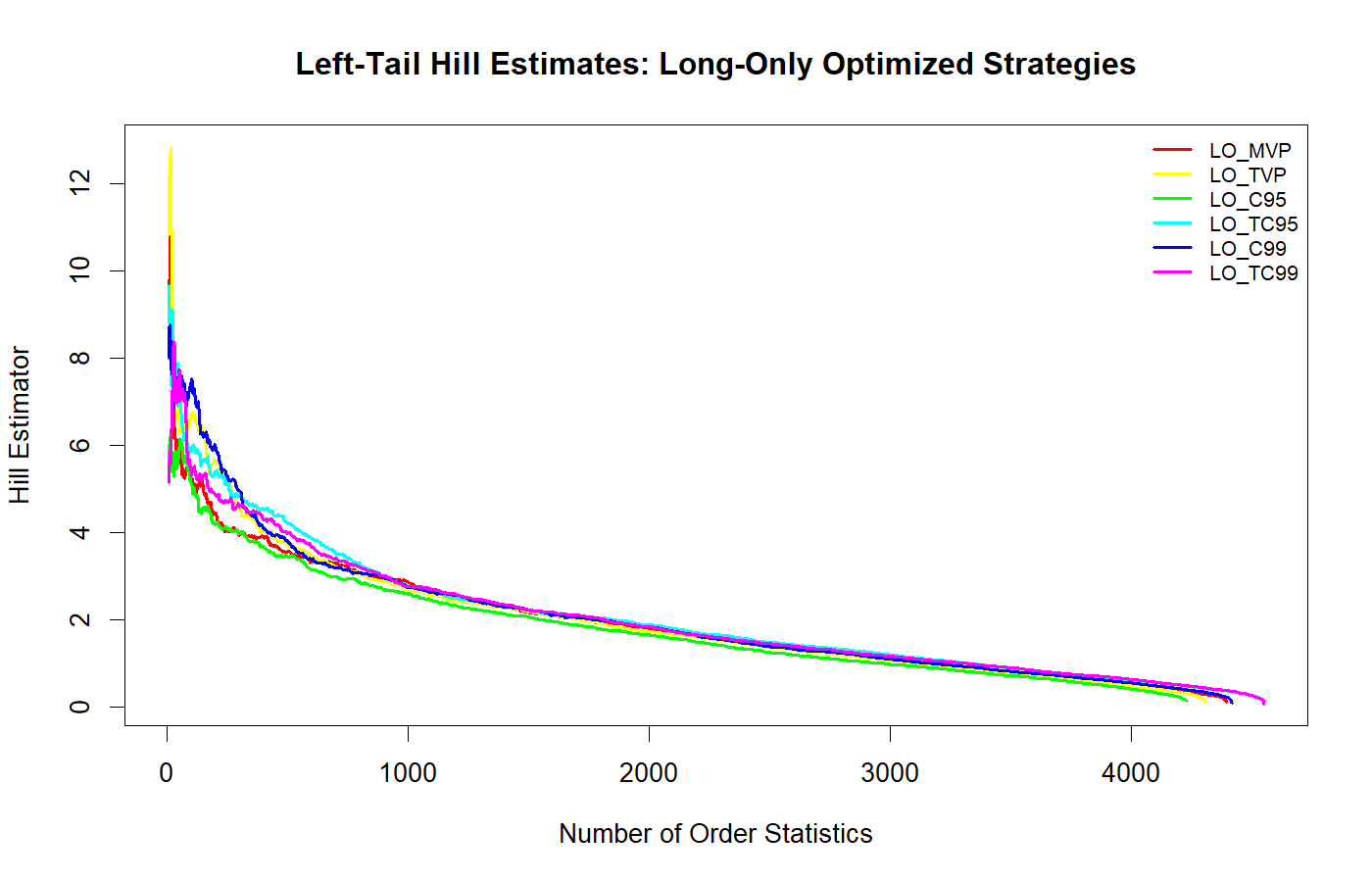}
\includegraphics[width=.49\textwidth]{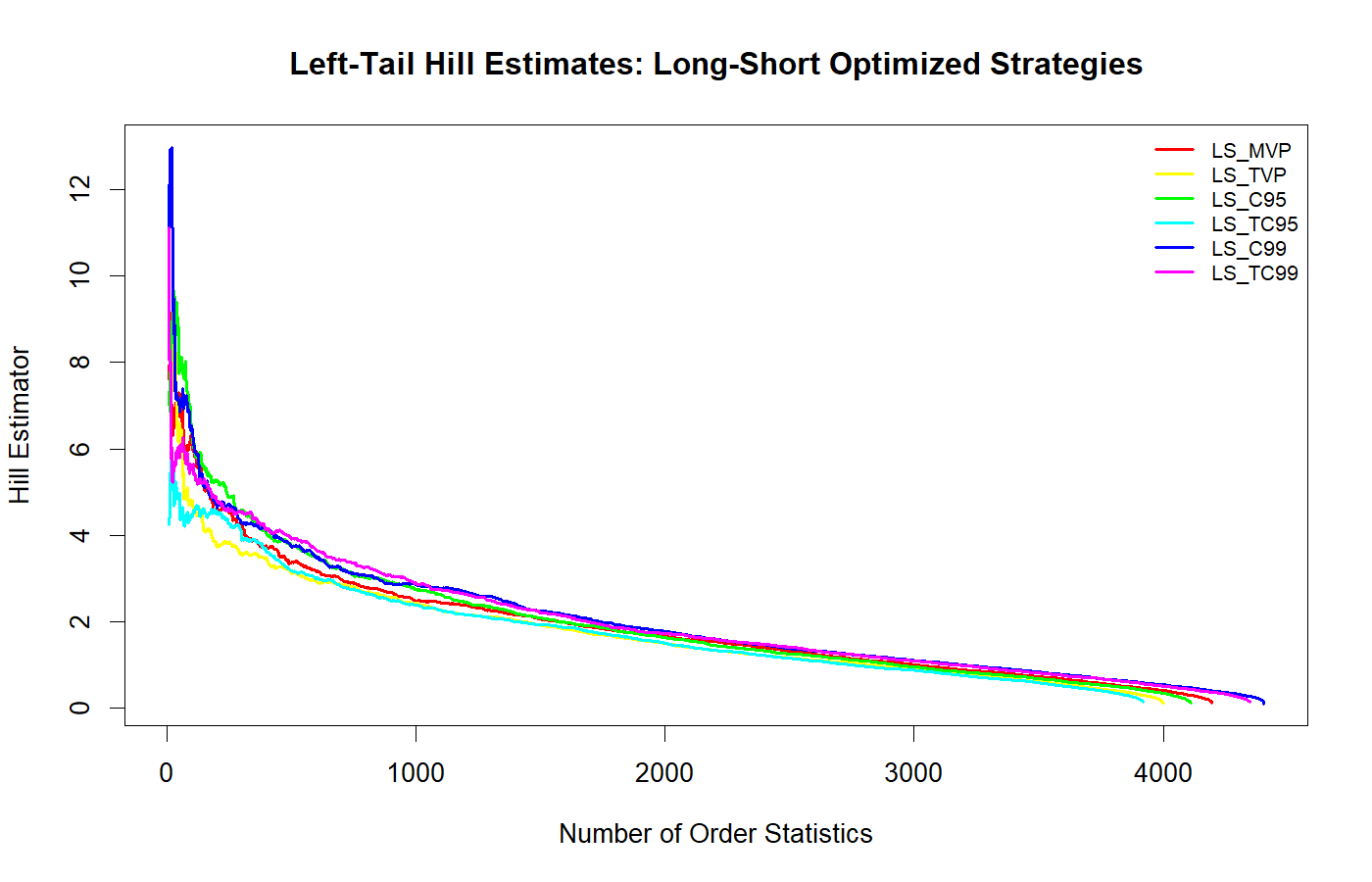}
\caption{Dynamic left-tail Hill estimators for long-only and long--short optimized strategies. Panel (a) reports dynamic long-only portfolios, while Panel (b) reports dynamic long--short portfolios.}
\label{figHillDynamicCombined}
\end{figure}

Figure~\ref{figHillDynamicCombined} reports the corresponding Hill estimates for the dynamic portfolios. Relative to the historical case, the dynamic curves remain heterogeneous at very low order statistics but become more tightly clustered as the threshold expands. In the dynamic long-only panel, C95 tends to lie below several other strategies over a substantial range of (k), suggesting relatively lighter downside-tail behavior. In the dynamic long--short panel, LS TVP and LS TC95 often remain near the lower portion of the Hill bundle, while LS C99 and LS TC99 more often appear among the upper curves.

\begin{figure}[h!]
\centering
\includegraphics[width=.8\textwidth]{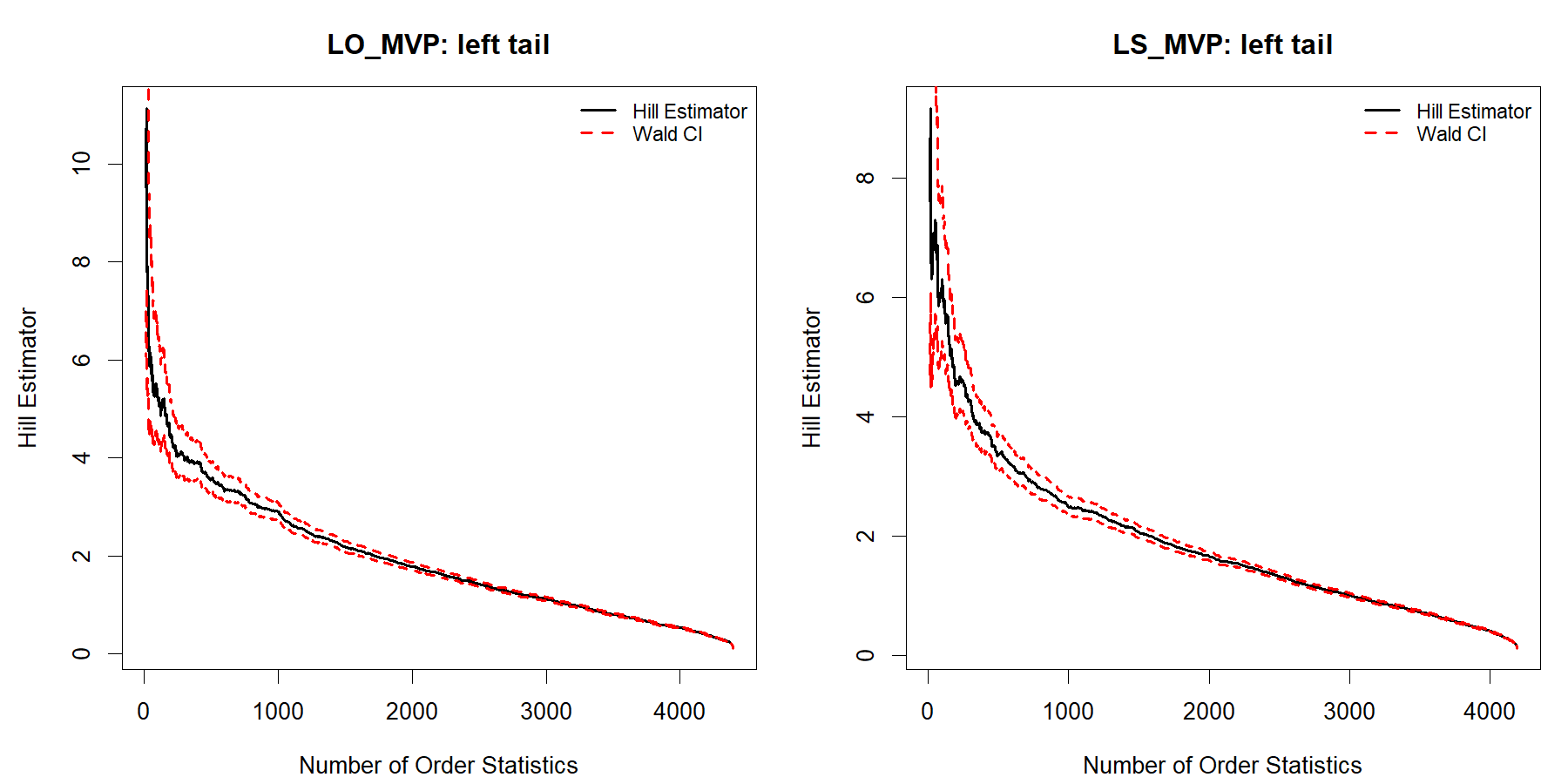}
\includegraphics[width=.8\textwidth]{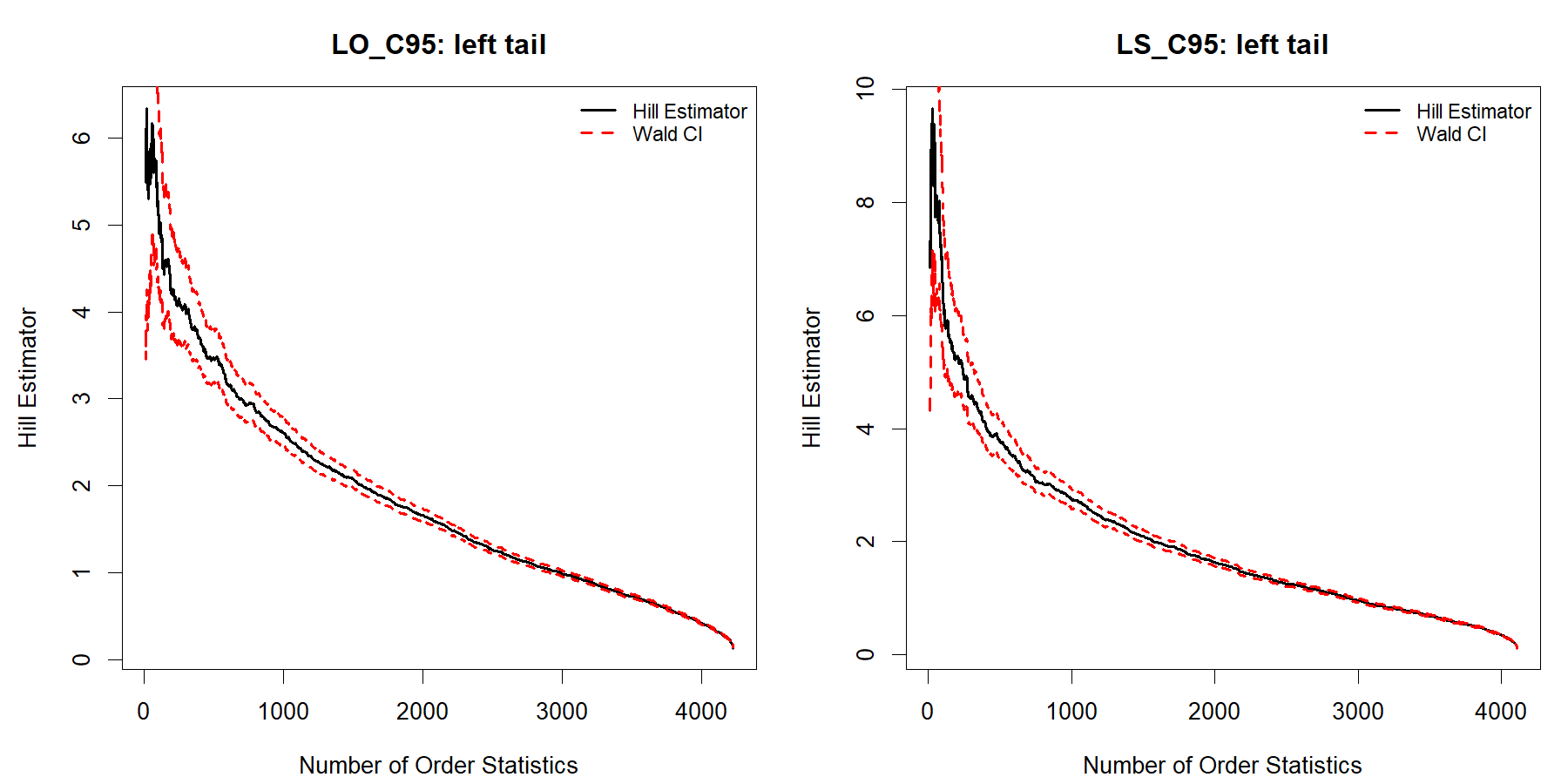}
\includegraphics[width=.8\textwidth]{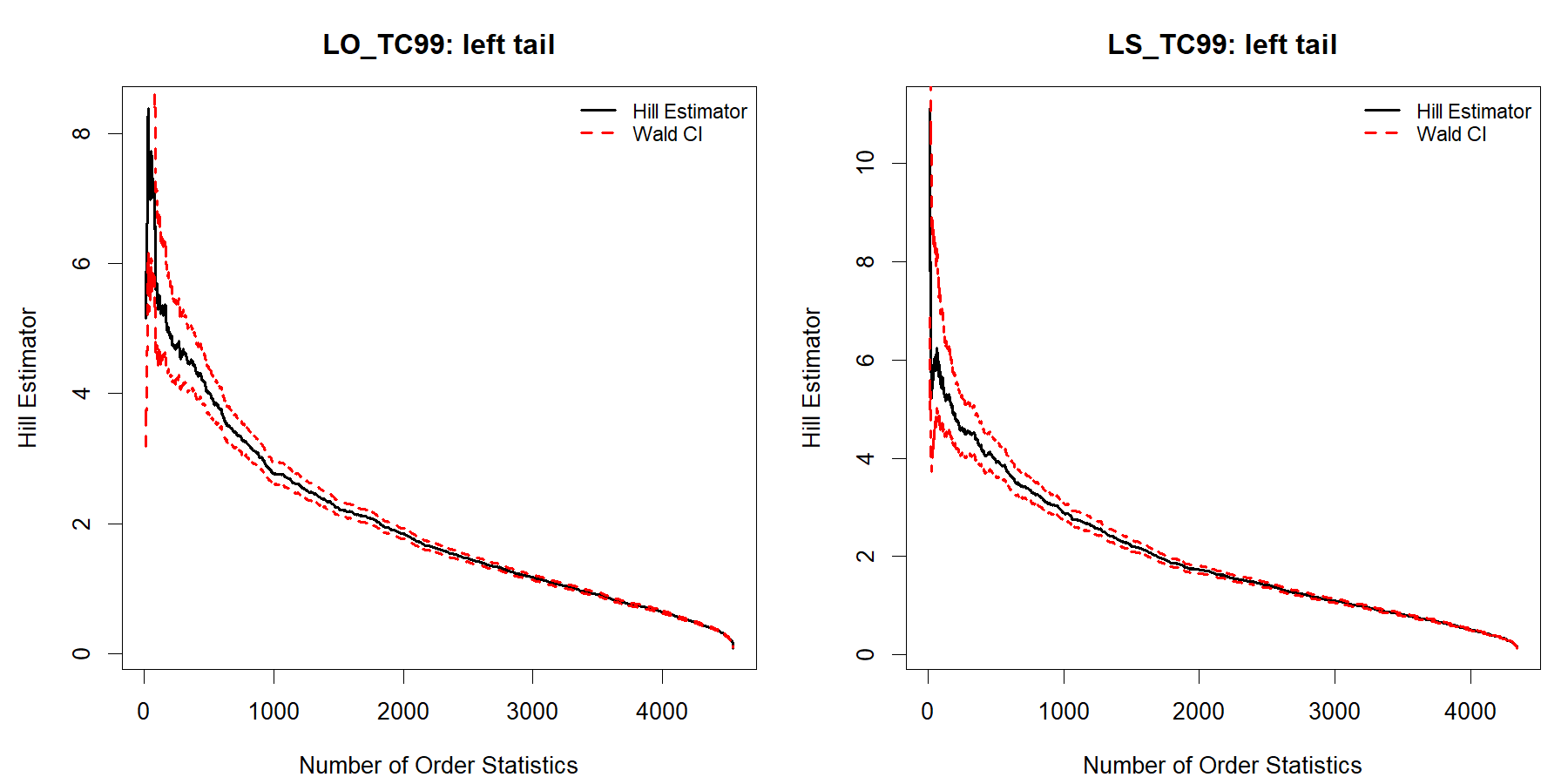}
\caption{Dynamic left-tail Hill comparisons for selected strategies. Panel (a) compares MVP, Panel (b) compares C95, and Panel (c) compares TC99.}
\label{figHillDynamicSelected}
\end{figure}

Figure~\ref{figHillDynamicSelected} provides dynamic long-only versus long--short comparisons for MVP, C95, and TC99. The MVP comparison suggests that limited shorting may improve downside-tail containment for the dynamic minimum-variance strategy. The C95 curves are closer, indicating that the CVaR-95 objective already imposes meaningful tail discipline. The TC99 comparison shows that long--short implementation can increase downside-tail sensitivity when combined with a far-tail, return-seeking objective.

Taken together, the Hill diagnostics show that portfolio optimization and dynamic rebalancing change the distribution of downside risk, but do not eliminate extreme-loss exposure. The effect depends on the risk measure, the constraint regime, and the degree of tail sensitivity embedded in the allocation rule.
\FloatBarrier
\subsection{POT--GPD Tail Diagnostics}
\label{subsecPOTGPD}

To examine the extreme tail more formally, we also implement a peak-over-threshold (POT) approach. The threshold (u) is set equal to the empirical 95th percentile of the loss distribution. For observations exceeding this threshold, the excess losses
\begin{equation}
Y=L-u
\label{eqPOTExcessLoss}
\end{equation}
are modeled using a Generalized Pareto Distribution (GPD). The resulting GPD shape parameter \(\xi\), scale parameter \(\beta\), threshold values, and exceedance counts are reported in Appendix~\ref{appPOTGPD}, Tables~\ref{tabHistoricalPOTGPD} and~\ref{tabDynamicPOTGPD}.

These estimates should be interpreted as diagnostic evidence rather than precise forecasts because the number of exceedances is limited in a daily sample of this length. Nevertheless, the POT--GPD results provide a useful robustness check on the empirical tail-risk evidence and connect the portfolio analysis to the broader extreme-value framework. Together with the VaR, ES/TCE, maximum drawdown, and Hill-estimator evidence, the POT--GPD diagnostics confirm that portfolio aggregation does not eliminate lower-tail risk and that tail exposure varies materially across optimization rules, constraint regimes, and dynamic rebalancing specifications.

\section{Discussion and Conclusion}
\label{secConclusion}

This study examined portfolio optimization and tail-risk analytics for a universe of actively managed investment funds composed predominantly of actively managed ETFs, with PTTRX retained as an actively managed fixed-income mutual-fund comparator. Using daily Bloomberg data from 4 December 2020 to 24 December 2025, the analysis compared buy-and-hold, mean--variance, CVaR-based, and tangency-type portfolio strategies under long-only and long--short constraints. The empirical framework combined historical full-sample optimization with a dynamic three-year rolling-window allocation design and evaluated the resulting portfolios using cumulative performance, risk-adjusted measures, downside-risk metrics, transaction-cost robustness checks, and extreme-tail diagnostics.

The results show that actively managed ETFs should be evaluated not only as stand-alone investment products, but also as interacting components within a broader portfolio opportunity set. The fund universe contains thematic equity, fixed-income, income-oriented, multi-asset, and alternative strategies with distinct return distributions, volatility profiles, drawdown behavior, and dependence structures. This heterogeneity creates meaningful diversification and allocation opportunities, but it also implies that portfolio outcomes depend on how funds are combined, constrained, and rebalanced through time.

The historical portfolio results indicate that tangency-type portfolios are generally the strongest competitors to the buy-and-hold benchmark in cumulative and risk-adjusted performance terms. This reflects their explicit emphasis on reward-to-risk efficiency. However, stronger upside-oriented performance can be accompanied by larger drawdowns and greater exposure to adverse tail outcomes. By contrast, minimum-variance and CVaR-minimizing portfolios behave more like defensive allocation rules. They tend to sacrifice upside participation, but provide more stable downside-risk profiles. The comparison highlights a central trade-off in active ETF portfolio construction: return-seeking strategies may improve conventional performance measures, while downside-risk strategies may be more appropriate for investors focused on capital preservation and tail-loss control.

The dynamic results show that adaptive rebalancing does not improve all strategies uniformly. In the long-only setting, the dynamic CVaR-95 portfolio is the most consistently attractive specification across several risk-adjusted criteria. In the long--short setting, dynamic tangency-CVaR portfolios perform strongly across multiple reward-to-risk measures, but their implementation is more demanding and their downside exposure can be larger. The transaction-cost robustness results further show that high-turnover dynamic strategies may experience substantial wealth drag once trading costs are incorporated. Thus, dynamic optimization is economically useful only when the gains from adaptability are large enough to compensate for turnover, estimation error, and implementation frictions.

The bootstrap Sharpe-ratio tests provide an important qualification. Across both historical and dynamic settings, Sharpe-ratio differences between competing mean--variance and CVaR-based strategies are generally not statistically significant at conventional levels. This result does not weaken the paper's contribution. Instead, it reinforces the need for a multidimensional evaluation framework. Portfolios that appear statistically similar under the Sharpe ratio may differ substantially in maximum drawdown, expected shortfall, lower-tail thickness, and POT--GPD tail diagnostics. The evidence therefore supports the use of CVaR, STARR, Rachev ratios, drawdown measures, Hill estimators, and peak-over-threshold diagnostics as complements to conventional mean--variance analysis.

The findings have practical implications for investors. Institutional investors, pension funds, endowments, wealth managers, and family offices may use actively managed ETFs as flexible building blocks for diversified allocation programs, but these products should be evaluated within a portfolio context rather than only through fund-level rankings or recent returns. Investors seeking stronger upside participation may prefer tangency-type allocations, while investors prioritizing drawdown mitigation and tail-risk control may find CVaR-minimizing portfolios more appropriate. Long--short strategies can expand the feasible allocation set, but their potential benefits must be weighed against financing costs, short-sale constraints, turnover, taxes, liquidity, and operational complexity.

Several limitations remain. The empirical results are based on a selected universe of 30 funds over a relatively recent sample period, and the active ETF market continues to evolve. Portfolio optimization is also sensitive to estimation error in expected returns, covariance matrices, and tail-risk measures, especially in dynamic and long--short implementations. Future research could extend this framework by incorporating transaction-cost-adjusted optimization, turnover penalties, liquidity constraints, robust covariance estimation, Bayesian or Black--Litterman inputs, and longer out-of-sample evaluation periods.

Overall, the evidence demonstrates that actively managed ETFs are best studied through an integrated portfolio-optimization and risk-management framework. Their economic value depends not only on individual fund performance, but also on diversification effects, allocation constraints, dynamic rebalancing, downside-risk control, and extreme-tail behavior. No single optimization rule dominates across all dimensions. The main implication is that active ETF portfolio construction requires a joint assessment of return performance, implementation feasibility, and tail-risk exposure.

\newpage
\bibliographystyle{apalike}
\bibliography{references_clean}

\newpage
\appendix
\appendix

\section{Additional Fund-Level Summary Statistics}
\label{appFundStats}

\begin{table}[h!]
\centering
\caption{Distributional Quantiles of Daily Log Returns}
\label{tabReturnQuantiles}
\scriptsize
\begin{tabular}{lrrrrrrr}
\toprule
Fund & Min & 1\% & 5\% & Median & 95\% & 99\% & Max \\
\midrule
ARKK & -0.0916 & -0.0609 & -0.0418 & 0.0017 & 0.0378 & 0.0524 & 0.1539 \\
ARKG & -0.0757 & -0.0620 & -0.0435 & -0.0006 & 0.0440 & 0.0686 & 0.1143 \\
ARKF & -0.0864 & -0.0536 & -0.0367 & 0.0022 & 0.0326 & 0.0466 & 0.1326 \\
ARKQ & -0.0784 & -0.0502 & -0.0331 & 0.0025 & 0.0303 & 0.0406 & 0.1314 \\
ARKW & -0.0960 & -0.0597 & -0.0383 & 0.0026 & 0.0349 & 0.0496 & 0.1282 \\
FBCG & -0.0799 & -0.0422 & -0.0239 & 0.0020 & 0.0215 & 0.0307 & 0.1241 \\
GQRE & -0.0451 & -0.0262 & -0.0143 & 0.0007 & 0.0132 & 0.0195 & 0.0556 \\
JEPI & -0.0572 & -0.0197 & -0.0097 & 0.0004 & 0.0087 & 0.0132 & 0.0691 \\
JPST & -0.0044 & -0.0043 & -0.0012 & 0.0002 & 0.0008 & 0.0010 & 0.0014 \\
JCPB & -0.0115 & -0.0088 & -0.0056 & 0.0001 & 0.0047 & 0.0064 & 0.0103 \\
DYNF & -0.0615 & -0.0300 & -0.0163 & 0.0014 & 0.0150 & 0.0245 & 0.0854 \\
PTTRX & -0.0138 & -0.0082 & -0.0058 & 0.0000 & 0.0047 & 0.0082 & 0.0107 \\
TOTL & -0.0234 & -0.0088 & -0.0053 & 0.0000 & 0.0045 & 0.0067 & 0.0242 \\
ANGL & -0.0260 & -0.0093 & -0.0060 & 0.0003 & 0.0048 & 0.0076 & 0.0289 \\
JMUB & -0.0132 & -0.0059 & -0.0037 & 0.0002 & 0.0027 & 0.0050 & 0.0086 \\
AOR & -0.0336 & -0.0156 & -0.0094 & 0.0007 & 0.0086 & 0.0123 & 0.0467 \\
AOA & -0.0473 & -0.0217 & -0.0112 & 0.0008 & 0.0103 & 0.0160 & 0.0625 \\
AOM & -0.0213 & -0.0120 & -0.0072 & 0.0005 & 0.0068 & 0.0090 & 0.0357 \\
AOK & -0.0186 & -0.0114 & -0.0058 & 0.0003 & 0.0056 & 0.0084 & 0.0238 \\
GAA & -0.0309 & -0.0175 & -0.0102 & 0.0007 & 0.0096 & 0.0131 & 0.0252 \\
KMLM & -0.0407 & -0.0176 & -0.0111 & 0.0000 & 0.0097 & 0.0140 & 0.0180 \\
DBMF & -0.0365 & -0.0197 & -0.0103 & 0.0007 & 0.0100 & 0.0169 & 0.0301 \\
QAI & -0.0399 & -0.0118 & -0.0058 & 0.0004 & 0.0055 & 0.0075 & 0.0311 \\
IVOL & -0.0375 & -0.0168 & -0.0089 & -0.0004 & 0.0096 & 0.0159 & 0.0324 \\
JAAA & -0.0104 & -0.0049 & -0.0035 & 0.0002 & 0.0012 & 0.0016 & 0.0056 \\
RIGS & -0.0301 & -0.0149 & -0.0091 & 0.0002 & 0.0089 & 0.0184 & 0.0261 \\
SRLN & -0.0132 & -0.0089 & -0.0058 & 0.0002 & 0.0022 & 0.0042 & 0.0212 \\
ELD & -0.0313 & -0.0179 & -0.0104 & -0.0001 & 0.0108 & 0.0205 & 0.0267 \\
MUNI & -0.0169 & -0.0065 & -0.0035 & 0.0002 & 0.0032 & 0.0051 & 0.0167 \\
PULS & -0.0050 & -0.0044 & -0.0010 & 0.0002 & 0.0006 & 0.0010 & 0.0014 \\
\bottomrule
\end{tabular}
\end{table}

The distributional quantiles provide additional evidence of heterogeneity across the actively managed fund universe. Equity-oriented and thematic funds exhibit substantially larger negative returns in the lower tail, while fixed-income and income-oriented funds generally display milder losses at the 1\% and 5\% quantiles. These differences support the use of CVaR-based optimization and extreme-value diagnostics in the main analysis.

\section{POT--GPD Tail Diagnostics}
\label{appPOTGPD}

The historical POT--GPD diagnostics show moderate variation in the estimated shape and scale parameters across portfolio rules and constraint regimes. Although the number of exceedances is limited, the estimates indicate that optimized portfolios retain nontrivial lower-tail exposure after aggregation. The long--short tangency-type portfolios generally show larger historical shape estimates than their long-only counterparts, suggesting that shorting flexibility may alter the structure of extreme downside losses.
\FloatBarrier
\begin{table}[h!]
\centering
\caption{POT--GPD Tail Diagnostics for Historical Portfolios}
\label{tabHistoricalPOTGPD}
\scriptsize
\begin{tabular}{llrrrr}
\toprule
Constraint & Portfolio & Threshold $u_{0.95}$ & Exceedances & Shape $\xi$ & Scale $\beta$ \\
\midrule
LO & MVP  & 0.812 & 25 & 0.122 & 0.315 \\
LO & TVP  & 0.905 & 25 & 0.091 & 0.398 \\
LO & C95  & 0.812 & 25 & 0.126 & 0.313 \\
LO & TC95 & 0.905 & 25 & 0.094 & 0.396 \\
LO & C99  & 0.826 & 25 & 0.141 & 0.306 \\
LO & TC99 & 0.905 & 25 & 0.096 & 0.395 \\
LS & MVP  & 0.806 & 25 & 0.092 & 0.328 \\
LS & TVP  & 0.901 & 25 & 0.183 & 0.345 \\
LS & C95  & 0.806 & 25 & 0.093 & 0.328 \\
LS & TC95 & 0.901 & 25 & 0.184 & 0.344 \\
LS & C99  & 0.826 & 25 & 0.141 & 0.306 \\
LS & TC99 & 0.901 & 25 & 0.181 & 0.346 \\
\bottomrule
\end{tabular}
\end{table}

\begin{table}[h!]
\centering
\caption{POT--GPD Tail Diagnostics for Dynamic Portfolios}
\label{tabDynamicPOTGPD}
\scriptsize
\begin{tabular}{llrrrr}
\toprule
Constraint & Portfolio & Threshold $u_{0.95}$ & Exceedances & Shape $\xi$ & Scale $\beta$ \\
\midrule
LO & MVP  & 0.178 & 26 & -1.477 & 0.378 \\
LO & TVP  & 2.563 & 26 & -0.409 & 1.775 \\
LO & C95  & 0.284 & 26 & -0.637 & 0.143 \\
LO & TC95 & 2.563 & 26 & -0.409 & 1.775 \\
LO & C99  & 0.199 & 26 & -1.210 & 0.378 \\
LO & TC99 & 2.563 & 26 & -0.409 & 1.775 \\
LS & MVP  & 0.175 & 26 & -1.847 & 0.477 \\
LS & TVP  & 2.562 & 26 & -0.458 & 2.044 \\
LS & C95  & 0.269 & 26 & -1.355 & 0.250 \\
LS & TC95 & 2.562 & 26 & -0.458 & 2.044 \\
LS & C99  & 0.210 & 26 & -1.169 & 0.315 \\
LS & TC99 & 2.562 & 26 & -0.458 & 2.044 \\
\bottomrule
\end{tabular}
\end{table}

The dynamic POT--GPD diagnostics reinforce the distinction between conservative and return-seeking dynamic strategies. Dynamic minimum-risk and CVaR-minimizing portfolios have much lower thresholds than the tangency-type portfolios, while dynamic tangency-type portfolios have larger scale estimates. This is consistent with the main text, where dynamic tangency-type portfolios display stronger upside potential but larger drawdowns and tail losses.
\FloatBarrier
\section{Additional Transaction-Cost Robustness}
\label{appTransactionCosts}

\begin{table}[h!]
\centering
\caption{Historical Portfolio Transaction-Cost Robustness at 50 Basis Points}
\label{tabHistoricalTCRobustness}
\scriptsize
\begin{tabular}{llrrrrr}
\toprule
Constraint & Strategy & Ann. Turnover (\%) & Gross TW & Net TW & Wealth Drag & Net Sharpe \\
\midrule
BHP & BHP & 0.00 & 118.42 & 118.42 & 0.000 & 0.507 \\
LO & MVP & 3.33 & 114.68 & 114.64 & 0.038 & 0.371 \\
LO & TVP & 3.33 & 119.18 & 119.15 & 0.039 & 0.542 \\
LO & C95 & 3.33 & 114.87 & 114.83 & 0.038 & 0.381 \\
LO & C99 & 3.33 & 114.99 & 114.95 & 0.038 & 0.385 \\
LS & MVP & 3.33 & 114.67 & 114.63 & 0.038 & 0.371 \\
LS & TVP & 3.33 & 118.44 & 118.40 & 0.039 & 0.517 \\
LS & C95 & 3.33 & 114.68 & 114.64 & 0.038 & 0.372 \\
LS & C99 & 3.33 & 114.96 & 114.93 & 0.038 & 0.383 \\
\bottomrule
\end{tabular}

\vspace{0.4em}
\parbox{0.94\textwidth}{\scriptsize \textit{Notes:} TW denotes terminal wealth from an initial value of 100. Net TW is computed after applying a 50 basis-point transaction cost per unit of turnover.}
\end{table}

The historical transaction-cost results show that the fixed-weight historical portfolios are only mildly affected by transaction costs because turnover is low. This contrasts with the dynamic transaction-cost results reported in the main text, where frequent rebalancing produces substantially larger wealth drag for several strategies.

\end{document}